\documentclass[twocolumn]{openjournal}

\usepackage[T1]{fontenc}
\usepackage{ae,aecompl}
\usepackage{natbib}
\usepackage{wasysym}

\usepackage{xcolor}
\begin{document}
\title{Dynamical Excitation as a probe of planetary origins}
\author{Brad M. S. Hansen}
\email[email]{ hansen@astro.ucla.edu}
\affiliation{ Mani L. Bhaumik Institute for Theoretical Physics,  Department of Physics \& Astronomy, University of California Los Angeles, \\ Los Angeles, CA 90095}
\author{Tze-Yeung Yu }
\affiliation{Mani L. Bhaumik Institute for Theoretical Physics,  Department of Physics \& Astronomy, University of California Los Angeles, \\ Los Angeles, CA 90095}
\author{Neel Nagarajan}
\affiliation{ Mani L. Bhaumik Institute for Theoretical Physics,  Department of Physics \& Astronomy, University of California Los Angeles, \\ Los Angeles, CA 90095}
\author{Yasuhiro Hasegawa}
\affiliation{Jet Propulsion Laboratory, California Institute of Technology, Pasadena, CA, 91109}

\begin{abstract}

We present a set of numerical simulations of the dynamical evolution of compact planetary systems migrating in a protoplanetary disk whose inner edge
is sculpted by the interaction with the stellar magnetic field, as described in \cite{YHH23}. We demonstrate that the resulting final distribution of neighbouring
planet period ratios contains only a small surviving fraction of resonant systems, in accordance with observations. The resulting planetary architectures are largely in
place by the end of the protoplanetary disk phase (within a few Myr), and do not require significant later dynamical evolution.

The  divergence of planetary pairs during gas disk dispersal also leads to the excitation of eccentricities when pairs cross mean motion resonances in a divergent fashion.
The resulting distribution of remnant free eccentricities is consistent with the values inferred from the observation of transit durations and transit timing variations. We furthermore
demonstrate that this conclusion is not significantly altered by tides, assuming standard values for tidal dissipation in Earth or Neptune-class planets. 

These results demonstrate that the observed spacing and residual dynamical excitation of compact planetary systems can be reproduced by migration
through a protoplanetary disk, as long as the inner disk boundary is modelled as a gradual rollover, instead of a sharp transition. Such an effect can be
achieved when the model accounts for the diffusion of the stellar magnetic field into the disk. The resulting divergence of planetary pairs during the
magnetospheric rebound phase breaks the resonant chains, resulting in a better match to observations than disk models with more traditional inner boundaries.

\end{abstract}

\maketitle


\section{Introduction}

The ongoing census of extrasolar planetary systems has revealed that the most common planetary systems orbiting Solar-like stars contain multiple planets
in the Earth--Neptune mass range, arrayed in compact configurations with orbital periods that frequently lie well within that of Mercury in our
Solar system \citep{Borucki10,Bat13,FTC13,PHM13,Cough16,Liss23}.
Despite their widespread occurrence, the origins of these planets is still
a matter of lively debate.

One school of thought suggests that such planets, especially those with volatile-rich envelopes, formed further from the star and migrated inwards due to interactions with the gaseous protoplanetary disk. The migration timescales associated with low mass planets are quite short, but gravitational interactions between multiple migrating planets can slow the evolution sufficiently to avoid losing the planets into the central star \citep{TP07,IL08,IL10,HN12,Coss14,CN14,Izi17}. The  observed high frequency of multiple planet systems also aligns nicely with this model. However, such models predict a high occurrence rate of mean motion resonances in such systems \cite{MS01,CN06,MW11}, which is not matched by observations \citep{Liss11,Fab14}. 

An alternative class of models suggests that the planets assembled in situ, with little radial migration \citep{HM12,CL13,BMF14,CT15,PNJ20}. These models can better match the observed spacing of planets in multiple systems \citep{HM13}, but require that the protoplanetary disks be either unusually massive or allow for significant inward migration of solid material as small bodies, prior to assembly. 

These two models represent two conceptual extremes, and recent work has explored scenarios which lie somewhat in between. The resonant chains produced by migration
may be broken by additional perturbations, such as from scattering by remnant planetesimals \citep{CF15,GC23,WML24}, late time mass loss \citep{MO20}  and excitation from distant giant planets \citep{LNV14,HA16,H17,HPD17,BA17,MDJ17,PL19,DNH19,PN20,PL21}. The simple resonant chain model may also be modified by more detailed migration models.
At late times, as the accretion rate drops, the magnetospheric cavity will expand, and the evolution of the torques during this `magnetospheric rebound' can destabilise some
resonances \citep{LOL17}.
 \cite{IBR21} consider a model in which migration occurs during the assembly process, with migrating embryos growing through the accretion of both planetesimals and planets.
 This leads to an inner planetary system composed of dense, resonant, chains of low mass planets that become dynamically unstable after the disk dissipates, leaving behind 
 final systems with a higher probability of being  non-resonant. \cite{HYH24} consider a model in which the migration occurs in a disk with an inner edge sculpted by the interaction with a stellar magnetic field, as described in \cite{YHH23}. During the magnetospheric rebound phase, as the disk dissipates, the evolution of this inner edge leads to resonant divergence  and, once again, disruption of most of the original chains.

A variety of statistics have been used to quantify the comparison between these models and observations, including the distribution of nearest neighbour period ratios, distribution of final planetary masses and orbital periods, relative frequency of multiple transit systems, and
 a variety of measures designed to quantify the statistical properties of planetary systems \citep{Laskar97,Chambers01,GF20,GB22,SSH24}.
 
Of particular interest is 
   the distribution of residual  orbital eccentricities.
   The most straightforward migration models predict very low eccentricities due to the eccentricity damping by  the gas disk. In the
case of in situ assembly models, very few systems are ever in resonance, and the eccentricity distribution is a remnant of the scattering during the assembly process \citep{HM13,DLC16}.  We expect the more sophisticated migration models to lie somewhere in between, because
 there needs to be a mechanism to break the chains after they have been formed, and the eccentricities will provide a probe of the resulting excitation \citep{Izi17,GB22}. Measuring
the level of residual dynamical excitation in planetary systems is therefore potentially a powerful discriminant between different origin scenarios. 

As a first step in this direction, we present here a full dynamical simulation of the migration scenario outlined in \cite{YHH23} and \cite{HYH24}. This describes migration of planets within a protoplanetary disk with a physically motivated inner edge, maintained by the interaction with the stellar magnetic field. The distinguishing feature of this model is the presence of a transition zone between the inner edge and a location where the disk torque reverses direction, so that planets pushed inwards of the torque reversal experience outwardly-directed migration. During the later stages of the disk lifetime -- as the magnetospheric cavity expands -- this leads to divergent evolution, and the breaking of many resonant chains. The result is a period-ratio distribution much more consistent with observations than other migration models,  and without invoking late-time dynamical instabilities.  Many pairs also experience temporary excitation of eccentricity during these late stages, when their divergent evolution passes through a mean motion resonance. This can potentially explain the puzzling persistence of small levels of free eccentricity in many of these systems \citep{HL14,HL17}, and so we wish to explore, in some detail, how the residual level of eccentricity in these systems can be used to probe planetary origins.

In \S~\ref{Sims} we describe our dynamical simulations of the planetary scenario described in \cite{YHH23} and \cite{HYH24}, and explore the dynamical mechanisms that
lead to the final levels of eccentricity excitation.  Tidal dissipation can reduce the observable eccentricity so, in \S~\ref{Tides} we explore the role played by tidal evolution in these systems.
 In \S~\ref{Excite} we compare these results to known observational constraints.

\section{Dynamical Model}
\label{Sims}

\cite{YHH23} presents a model of a protoplanetary disk, incorporating both viscous and radiative heating, in which low mass planets migrate inwards due to
the Type I migration torque formulae from \cite{HN12}. The models also incorporate a disk inner edge sculpted by the diffusion of the stellar magnetic field into
the disk. The effective viscosity near the inner edge is increased as a result of the stellar field torques, resulting in a peak of the surface density at a location
exterior to the inner edge (the nominal magnetospheric cavity). This causes a reversal of the migration torque at the surface density peak, and a change in the direction of planetary
migration before the planet reaches the true disk inner edge. The resulting influence of this torque behaviour, and its evolution over time, was studied in \cite{HYH24}.  Most simulated 
systems  initially formed a chain of resonant pairs during inward migration -- a result consistent with other migration models. However, as the accretion rate
drops, the location of the torque reversal moves outwards during the magnetospheric rebound phase. At this point, many of the migrated planets lay interior
to the torque reversal, and so many planet pairs experienced divergent migration at late times. This resulted in most of the original resonant chains
being  broken during the dissolution of the protoplanetary disk.

The simulations of \cite{HYH24} were based on a semi-analytic model that included only the three lowest first order resonances. 
This was sufficient to demonstrate the most important feature of the model -- namely that it produced a population of planets with a low resonant fraction, as observed.
However, our goal in this paper is  to extend this study to quantify the level of residual dynamical excitation in these model systems. To do this properly,  we wish to include the full range of gravitational interactions between
the planets in each system. Therefore, we have simulated planetary migration and dynamical evolution  following the same setup as in \cite{HYH24}, but now
 using the {\it Mercury} integrator \citep{Chambers99} to follow the full dynamical evolution.
In addition to the direct calculation of gravitational interaction between all bodies in the system, the planets are subject to the same gaseous torques as
described in \cite{HYH24}, following the underlying disk model. We use the same disk torques model described in equations (1)--(4) in \cite{HYH24}, which is
based on the $B_0=1.7$kG and $n=3.5$ model from \cite{YHH23}.

For the selection of planetary initial conditions, we follow the model of \S~5.2 of \cite{HYH24}. 
Masses for each system represent a particular Kepler analogue, with a mass model based on the observed radii and the mass--radius relationship from
\cite{CK18}.
This approach means that our model should contain the same range of masses and neighbor mass ratios as the observed Kepler systems (although
the final period ratios may differ from the corresponding observed system, depending on the migration model and the probabilistic nature of resonant capture).
The initial semi-major axis for the innermost planet is 0.08~AU and the starting semi-major axis for each additional planet is a factor 1.5 larger than the 
inner neighbour. This ensures that all planet pairs start with period ratios between the 3:2 and 2:1 resonances. It is likely that the planets assembled further 
out and migrated inwards, but \cite{HYH24} demonstrated that the memories of the initial conditions are lost once the system enters a resonance lock, so we have
focussed our computational resources on the approach to resonance capture and later divergence.
Initial eccentricities are set to 0.1 and
initial inclinations are chosen with a dispersion of $0.1^{\circ}$. The initial values of the argument of perihelion, the line of nodes, and the mean anomaly are all
given a random value between 0 and 360$^{\circ}$. As a result of the dissipative forces we have added to the dynamics, we use the Bulirsch-Stoer integrator
in {\it Mercury}, with a default timestep of 6 hours (because many of our systems ultimately evolve to short orbital periods).

\subsection{Two Planet Systems}
\label{PairSection}

\begin{figure}
\centering
\includegraphics[height=4.0cm,width=4.0cm,angle=0,scale=2.2]{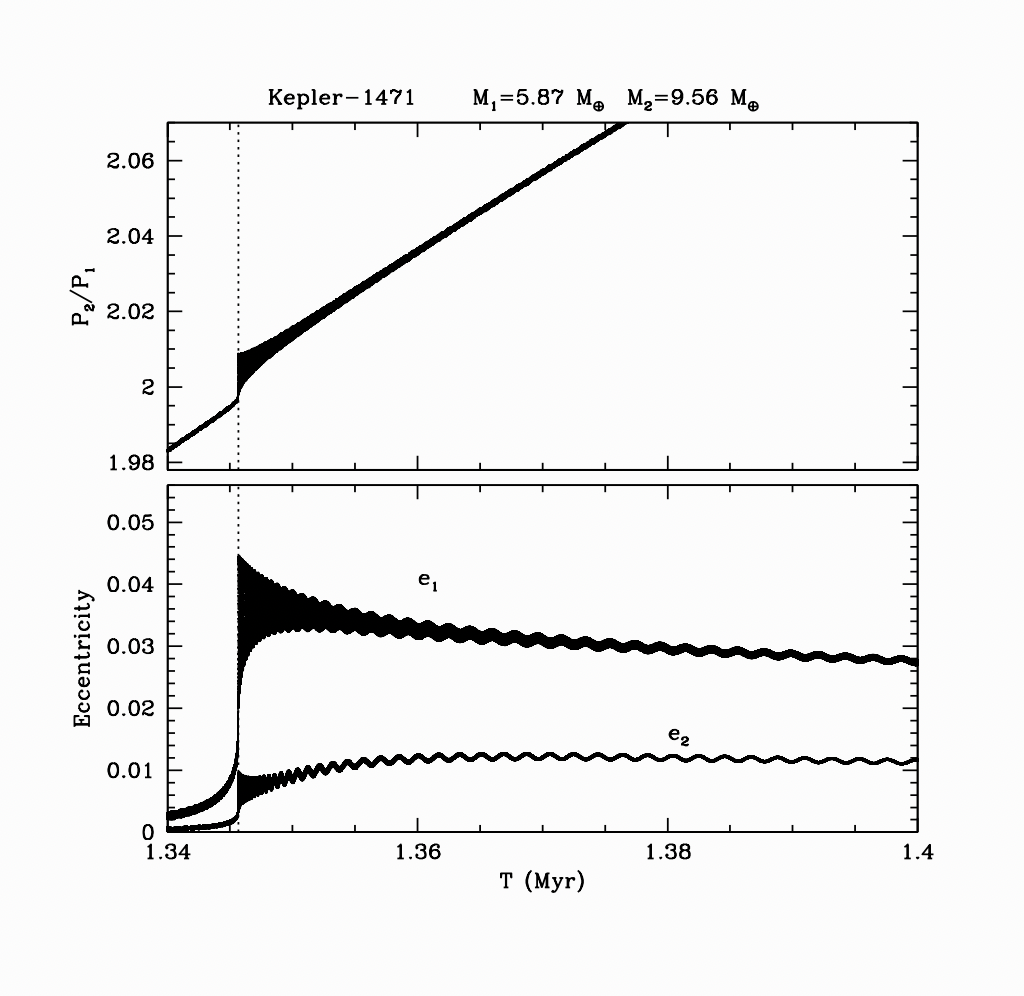}
\caption{The upper panel shows the evolution of the period ratio of the planetary pair in the system modelled after Kepler-1471. The jump at age 1.345~Myr is the
divergence crossing of the 2:1 resonance. In the lower panel, we see a corresponding spike in the eccentricity of both planets, most of which remains after the
disk has finally dispersed.
 \label{PairJump}}
\end{figure}

The migration of individual planets in this model are driven entirely by the gas torques and closely follow the same evolutionary path as described in \cite{YHH23}.
The evolution of two planet systems does incorporate mutual gravitational interactions, and we have simulated planetary pairs with masses chosen using the
above procedure.  Once again, the semi-analytic models, in this case, \cite{HYH24} capture the dynamics accurately. Most importantly, for the purposes of this
paper, we observe eccentricity excitation when pairs cross a mean motion resonance in the latter stages of the disk evolution. At late times, during the dispersal of the disk,
the torque reversal location is moving outwards, and many planetary pairs are diverging. The crossing of a resonance (in most cases studied here, the 2:1 mean motion
resonance) results in a transient eccentricity excitation. Once the pair has passed through the resonance, the eccentricities begin to damp again. However, many of these
divergent crossings occur late in the disk dispersal and the damping is incomplete, leaving a residual eccentricity in the system.

Figure~\ref{PairJump} shows an example of this, that occurs in our analog of the Kepler-1471 system. The inner planet has a mass of $5.87 M_{\oplus}$
and the outer planet has a mass of $9.56 M_{\oplus}$. This pair initially captures into the 3:2 mean motion resonance, during its inward migration. Since, the outer planet is more massive, it remains coupled to the disk for longer  during the magnetospheric rebound phase, and the
resulting diverging pair eventually crosses the 2:1 resonance at 1.34~Myr. Both planets receive a spike in the eccentricity during this passage. The pair
eventually freezes out at a period ratio $P_2/P_1=2.28$, with eccentricities $e_1=0.022$ and $e_2=0.009$.  As noted by \cite{LDH24}, such divergent
resonance crossings have the
potential to dictate the final residual eccentricities exhibited by transiting planets. We will examine the population-level consequences of this  in \S~\ref{Exciting}. 

\begin{figure}
\centering
\includegraphics[height=4.0cm,width=4.0cm,angle=0,scale=2.2]{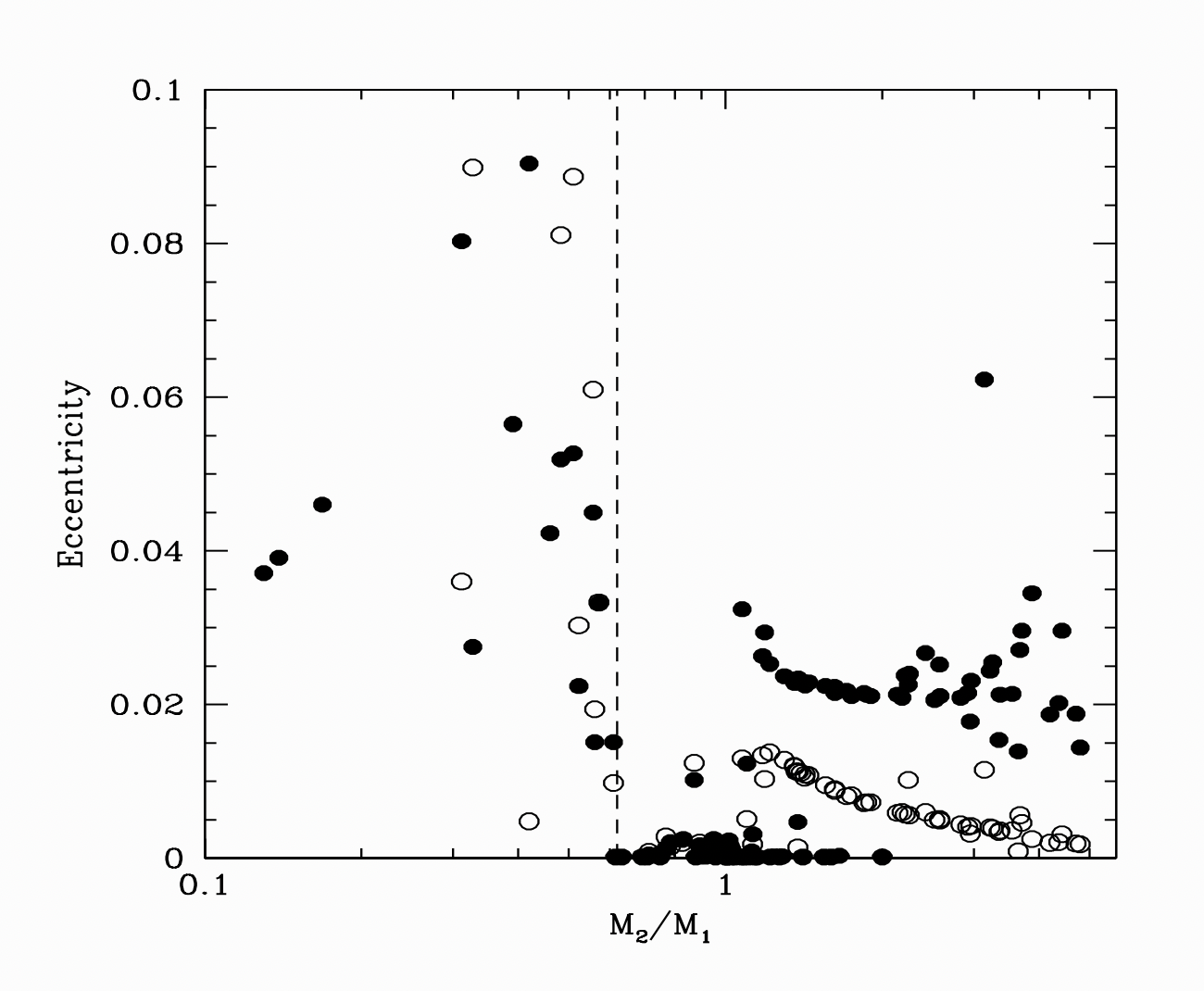}
\caption{The solid points show the inner planet eccentricity after 2~Myr, while the open circles show the outer planet eccentricity. The distribution of eccentricities
as a function of planetary mass ratio breaks up into three regimes. To the left of the vertical dashed line, i.e. at 
  small $M_2/M_1$, all systems remain in resonance to the end of the simulation. Thus,  the eccentricities are determined by the resonant lock of the pair.
 For mass ratios around unity, the final eccentricities are small, because the pairs break out of resonance and damp eccentricity.
On the right, the eccentricities are large again because the pairs diverge far enough to cross another mean motion resonance, resulting in late-time eccentricity excitation that persists to the end of the simulation.
\label{PairMe}}
\end{figure}

The final level of eccentricity excitation is a function of the planetary mass ratio, as is shown in Figure~\ref{PairMe}.  Systems in which the planets have mass ratios
close to or slightly below unity do not show a large amount of eccentricity. This is because both planets decouple from the disk at about the same time, and so there
is not sufficient differential movement in semi-major axis to cause the system to break the original resonance lock and then evolve to the point where they cross another
resonance. For systems with $M_2>M_1$, the outer planet moves far enough to generate such a divergent crossing (as in Figure~\ref{PairJump}), resulting in the
remnant eccentricities shown on the right of Figure~\ref{PairMe}. For systems with $M_2/M_1<0.6$,  it is the inner planet that remains coupled to the disk for longer
and so the outer planet will get swept along and remain in resonance. The larger eccentricities on the left of the vertical dashed line in Figure~\ref{PairMe} are therefore a consequence of maintaining
the resonant lock to the end of the simulation.

\subsection{Triple Planet Systems}

The two planet systems follow the semi-analytic evolution of \cite{HYH24} quite closely. However, systems with more planets can potentially introduce significantly
more dynamical complexity. The migration rate is a function of both mass and semi-major axis, so individual pairs may converge or diverge, depending on their parameters.
The interaction between these pairs can introduce effects not treated in the semi-analytic models.

\begin{figure}
\centering
\includegraphics[height=4.0cm,width=4.0cm,angle=0,scale=2.2]{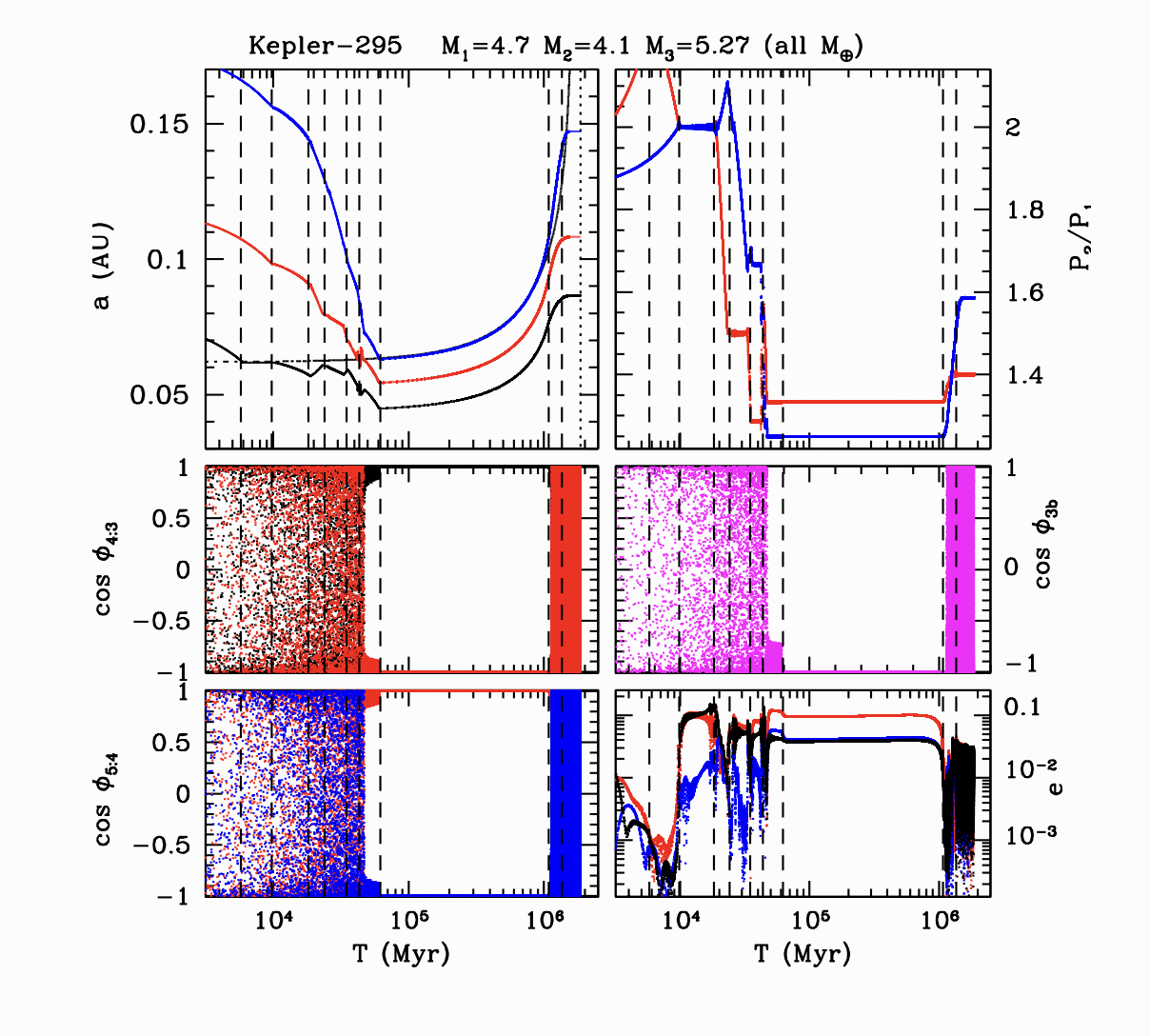}
\caption{The upper left plot shows the evolution of the semi-major axes for a three planet system under the influence of the torques derived from the protoplanetary
disk. The black curve represents the innermost planet, the red curve the middle planet and the blue curve the outer planet.  The dotted line tracks the location of the torque reversal.  Important transitions in the system
properties are marked by vertical dashed lines.   The right-hand plot shows the evolution of the nearest neighbour period ratios (color coded by the outer planet in each pair). The panel in the lower right shows the evolution of the eccentricity of each planet. The remaining panels show the evolution
of different resonant angles. The two on the left show the two-body resonant angles for the outer pair (4:3) and inner pair (5:4), color-coded by the planetary
precession term in the resonant argument. The magenta curve on the right shows the evolution of the three-body resonant angle.The system masses are in units of $M_{\oplus}$.
 \label{Kep295}}
\end{figure}

An example is shown in Figure~\ref{Kep295}, a system with masses motivated by the triple planet system Kepler-295. As in the semi-analytic
models, the innermost planet is the first to encounter the torque reversal and therefore halt its inward migration. The second and third planets quickly
lock into a 2:1+2:1 chain of mean motion resonances, driving the innermost planet interior to the torque reversal location. However, the disk torques are
strong enough to drive the planets closer together, resulting in a cascade down through resonances until a 4:3+5:4 chain is created, which reaches a
relatively stable configuration once the third planet reaches the torque reversal. The system remains in this resonant configuration as the torque reversal
begins to move outwards. Although all three planets are, at this point, migrating outwards rather than inwards, their resonant lock is maintained by the fact that
their relative outward migration rates still yield convergent evolution.

 At late times, the outward evolution accelerates  as the disk surface density drops, and the planetary migration rates cannot keep up. As the
 planet migration decouples from the disk evolution, the pairs can no longer maintain the converging relative motion, and the pairs drop out of resonance. The
resonant chains are broken, so 
that the final period ratios are not in resonance (nor is the trio in a three-body resonance). 

As long as the pairs remain in resonance, the eccentricities are maintained by the resonant lock \citep{TP19} in the face of the disk eccentricity
damping. The lower right hand panel of Figure~\ref{Kep295} shows that the equilibrium eccentricity values shift at each change in the
resonance configuration, as the signs of the torques driving the migration change.
This holds even during the outward migration, as long as the pairs are still converging, but the eccentricities drop rapidly once
the pairs leave resonance. Figure~\ref{Kep295} also shows that the divergence can be far enough to allow pairs to cross multiple resonances in
a divergent fashion. In the case shown here, 
 we see that the 
 outer pair diverges  from the original 5:4 commensurability and passes through the 4:3 and  3:2 resonances at late times, resulting
in a spike in the eccentricity during each divergent crossing. This can be observed in the lower right panel of Figure~\ref{Kep295}, and leaves the system
with a low level of finite eccentricity at late times. This is an essential contribution to the final level of dynamical excitation, and will be explored
further in \S~\ref{Exciting}.

\begin{figure}
\centering
\includegraphics[height=4.0cm,width=4.0cm,angle=0,scale=2.2]{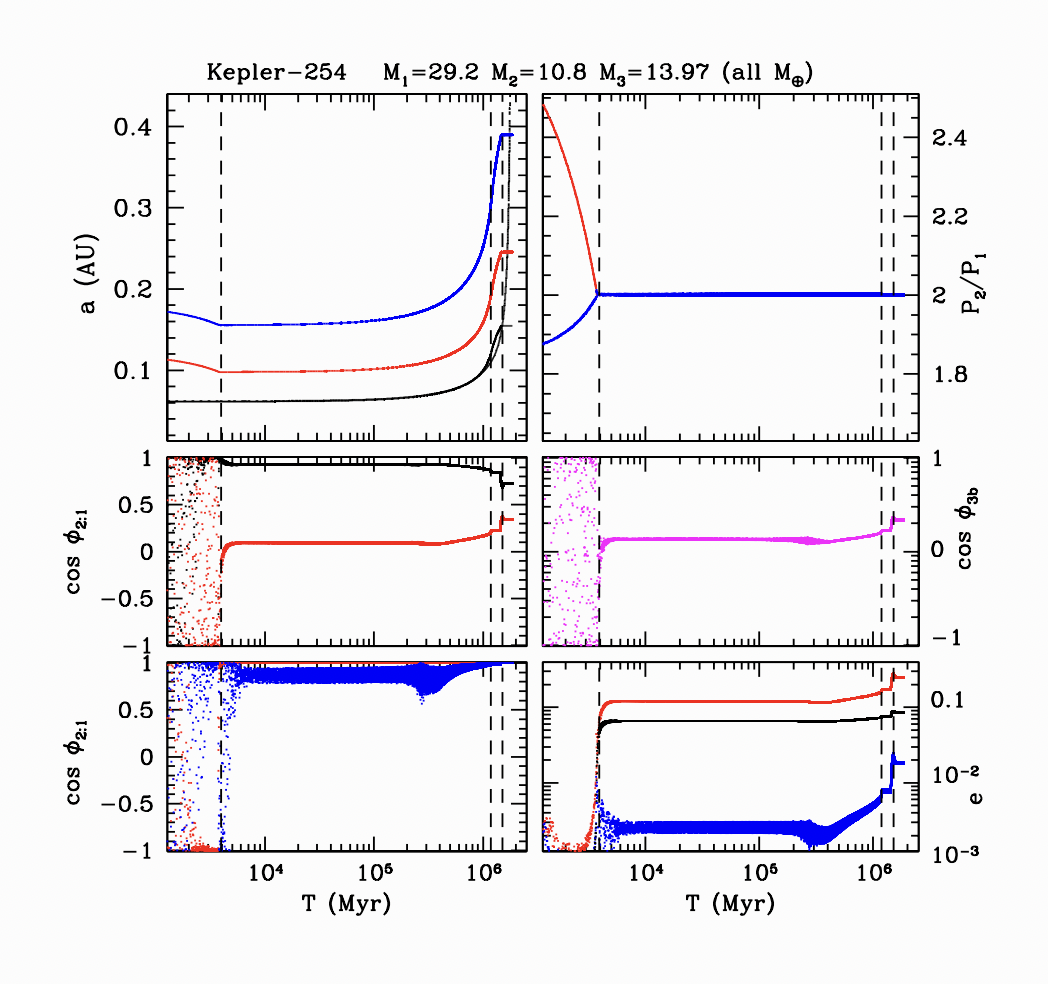}
\caption{The upper left plot shows the evolution of the semi-major axes for a three planet system under the influence of the torques derived from the protoplanetary
disk. The black curve represents the innermost planet, the red curve the middle planet and the blue curve the outer planet. Important transitions in the system
properties are marked by vertical dashed lines.  The dotted line tracks the location of the torque reversal. The right-hand plot shows the evolution of the nearest neighbour period ratios (color coded by the outer planet in each pair). The panel in the lower right shows the evolution of the eccentricity of each planet. The remaining panels show the evolution
of different resonant angles. The two on the left show the two-body resonant angles for the outer pair (2:1) and inner pair (2:1), color-coded by the planetary
precession term in the resonant argument. The magenta curve on the right shows the evolution of the three-body resonant angle.The system masses are in units of $M_{\oplus}$.
 \label{Kep254}}
\end{figure}

However, not every system breaks the resonant chain. Figure~\ref{Kep254} shows the evolution of a three planet system whose masses are based on
those of the triple planet Kepler-254 system, and which remains in the resonant chain after
the end of the disk phase. This is one of the more massive systems in our simulation set, with all the planets $> 10 M_{\oplus}$. The strength of the disk torque
scales with mass, so more massive planets remain coupled to the gas disk for longer. Furthermore, the most massive planet in this system is the innermost one, so it
will continue to move outwards longest, and maintain the convergent nature of the innermost pair, even as both planets move outwards. 

A further interesting feature of this system is that the resonant angles of the inner pair do not librate about 0 and $\pi$, but rather about some intermediate values.
These are examples of the `asymmetric equilibria'  of three-body
resonant chains, discussed by
\cite{WZB24} (see also \cite{Del17}). They can be heuristically understood as a consequence that planets in a chain of resonances are subjected to driving from multiple oscillators,
leading to equilibria that cannot be found in the two planet case. It is worth noting that the corresponding three-body resonant angle also has a different
libration center. As the disk dissipates and the driving torques decrease, these offsets are reduced, but they retain a non-zero difference at the end of the evolution.

\subsection{Higher Multiplicity Systems}

\begin{figure}
\centering
\includegraphics[height=4.0cm,width=4.0cm,angle=0,scale=2.2]{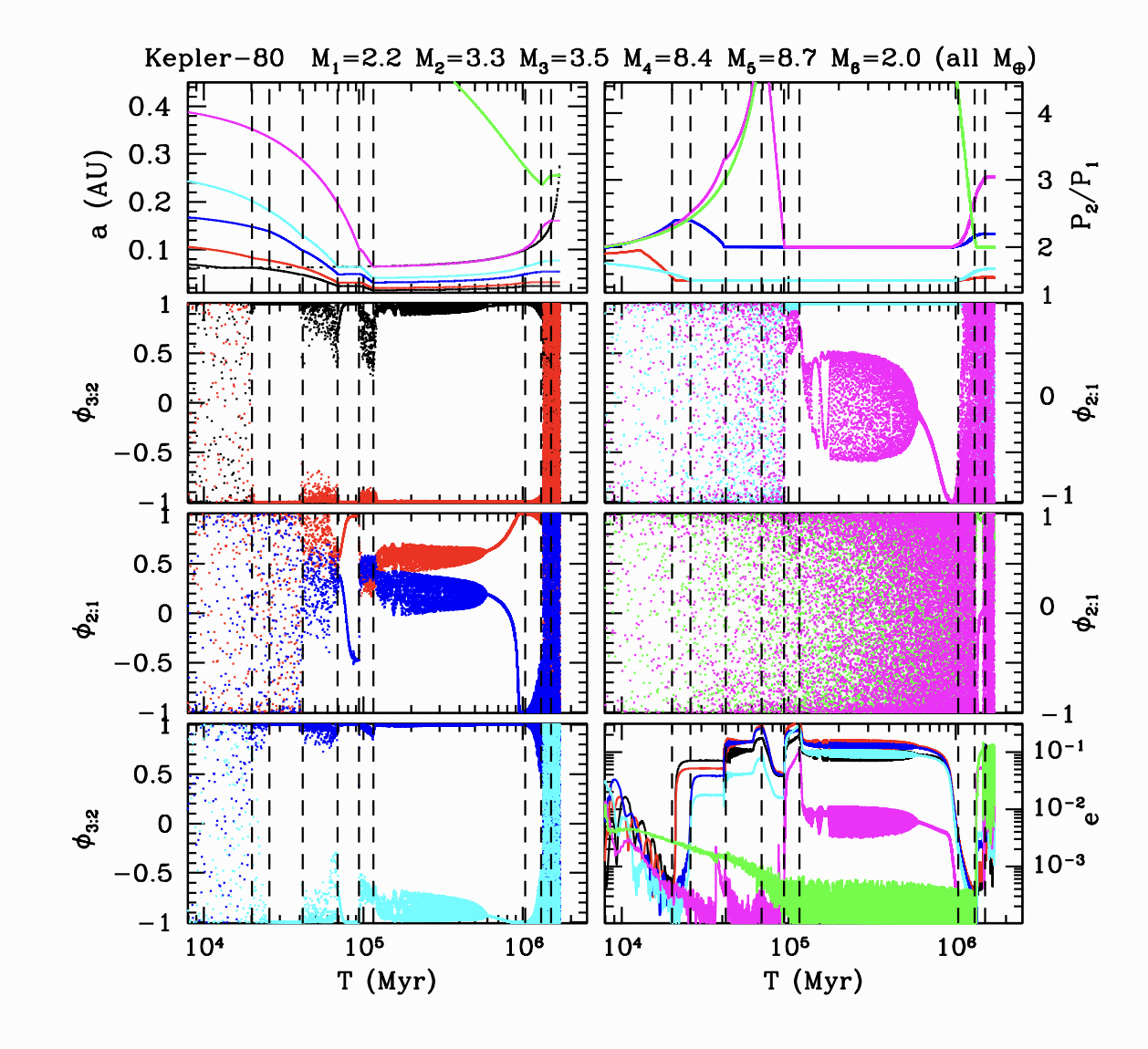}
\caption{The upper panels show the evolution of the semi-major axes (left) and nearest neighbor period ratios (right) for a six planet migrating system.
The panel in the lower right shows the corresponding evolution of the eccentricities. The other panels show the cosines of the most important resonant
angle for each of the 5 nearest neighbour pairs, with the colours corresponding to the planets in the upper left panel. As before, the dotted line in the upper
left shows the location of the torque reversal, and the vertical dashed lines indicate important epochs (when resonances are established or broken).
 \label{Kep80}}
\end{figure}

Another advantage of a full N-body simulation is that one can include an arbitrarily large number of planets. As such, in addition to the three  planet systems simulated in
\cite{HYH24}, we simulated mass distributions corresponding the
the Kepler  quadruple, quintuple and sextuple systems.  This followed the same procedure as in \cite{HYH24}, wherein the masses are a chosen from the probability
distribution of \cite{CK18}, estimated from the observed planetary radii, and we exclude systems with Jupiter-class ($R>6 R_{\oplus}$) planets.
Figure~\ref{Kep80} shows the evolution of a system whose masses were drawn from the distributions
for the Kepler-80 system. The inner five planets of this system sequentially capture into a resonance chain of 3:2+2:1+3:2+2:1 over the first $10^5$~years. This
chain remains locked until an age $\sim 10^6$~years, before it too begins to fall apart as the planetary migration falls behind the disk evolution. The resonant
angles in this chain also show a variety of symmetric and asymmetric equilibria. 
The first and third pairs librate in a traditional configuration,
but the second and fourth pairs show libration about asymmetric equilibria. As the disk torques maintaining the resonant lock weaken, these equilibria evolve
to more traditional libration centers, just before the resonances break and the eccentricities damp.

The outermost planet in the system takes a long time to migrate inwards (migration rates are slower further from the star) and only locks into the
2:1 resonance at late times (when the other planets have already broken their resonant locks). This lock is relatively brief, but it does leave some
residual eccentricity excitation in the outermost planet. The inner planets are not strongly coupled to it and so receive only a limited eccentricity excitation.

\subsection{Dynamical Instability}

\begin{figure}
\centering
\includegraphics[height=4.0cm,width=4.0cm,angle=0,scale=2.2]{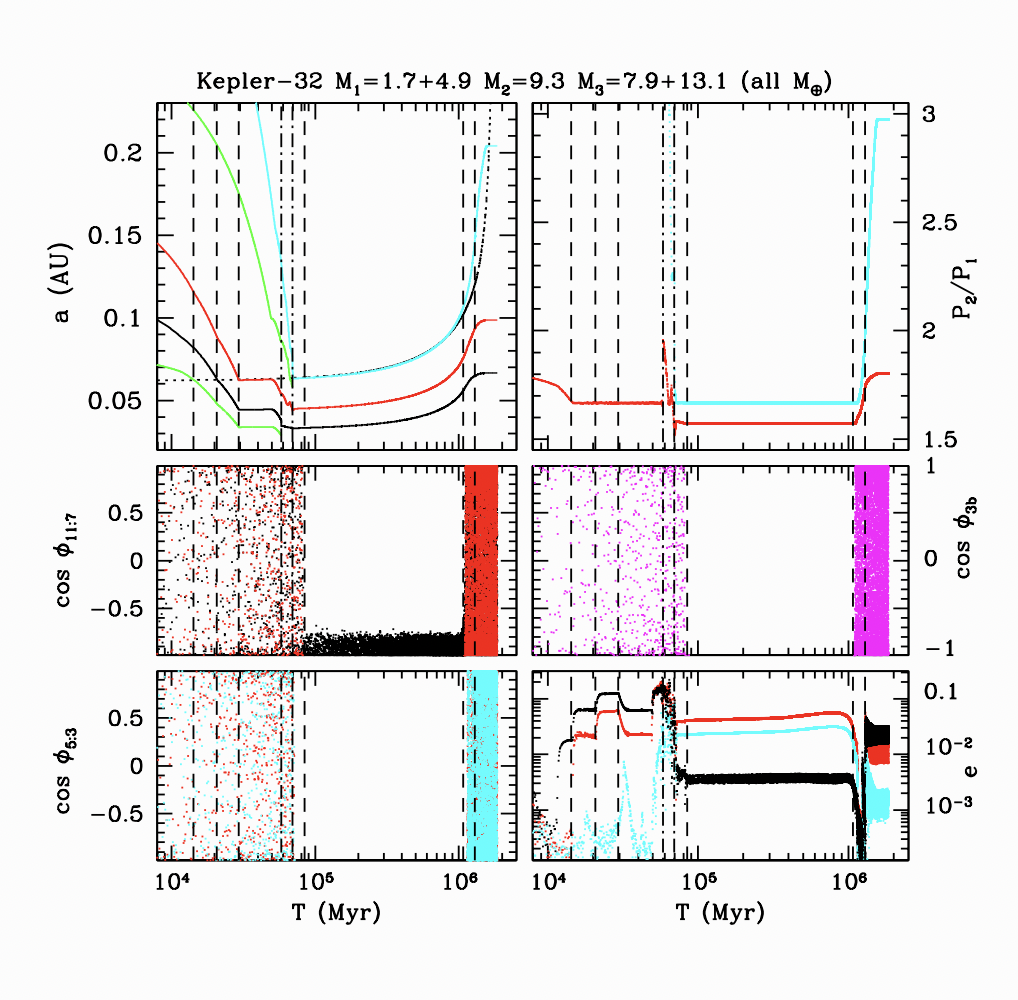}
\caption{The upper panels show the evolution of the semi-major axes (left) and nearest neighbor period ratios (right) for a  migrating system of (initially) five
planets.  In the upper left panel, the two green trajectories represent the two planets that are lost to merger with others. In the other panels, only trajectories
of surviving planets are shown, coloured according to our previous convention, as applied to planets at the end.
The panel in the lower right shows the  evolution of the eccentricities. The other panels show the cosines of the most important resonant
angle for each of the 5 nearest neighbour pairs, with the colours corresponding to the planets in the upper left panel. As before, the dotted line in the upper
left shows the location of the torque reversal, and the vertical dashed lines indicate important epochs (when resonances are established or broken).
 \label{Kep32}}
\end{figure}

The bulk of our simulated systems evolve through the migration and rebound phase without undergoing any dynamical instabilities, but some
do experience sufficient eccentricity excitation that some planets cross orbits and collide. We evolve these systems with {\em Mercury}, assuming perfect merging.
An example of such a system is shown in Figure~\ref{Kep32}.

This shows a system of, initially, five planets, with masses chosen to represent Kepler-32. Figure~\ref{Kep32} shows the five planet system evolving inwards
in the standard fashion (with green curves showing the two planets that are eventually merged with others). The inner three planets settle into a stable configuration
that tracks the torque reversal, while the outer two planets are still migrating inwards. When the fourth planet enters the resonant lock, it drives the interior planets
inward, excites eccentricity and generates instability. This results in the merger of the inner pair ($\sim 6 \times 10^4$ years). The continued inward migration
of the outermost planet (also the most massive in the system) soon results in an additional merger of the outer two planets (at $\sim 7 \times 10^4$ years). 

The end result of this is a three planet system following the same outward evolution as we see in other systems, except for the unusual resonant
configuration of 11:7+5:3. This also yields an unusual three-body resonant configuration of $10 n_3-17 n_2+7 n_1 \sim 0$, as shown in Figure~\ref{Kep32}.
As in other systems, this exotic resonant configuration is eventually disrupted by the outwards evolution, and the planets experience a similar late
eccentricity kick due to divergent crossing of the 2:1 resonance at late times. 

The overall fraction of systems that undergo dynamical instability increases with the multiplicity of the system. Of the original 107 triple systems simulated,
8 experienced dynamical instability and a planetary collision (a fraction of 7.5\%). Of the original 39 quadruple systems, 11 experienced instability (28\%),
while 12/18 (67\%) of high multiplicity systems experienced instability. Overall, 31/164 (19\%) of simulated systems experienced instability.

\subsection{Summary Statistics}
\label{SumStat}

\begin{figure}
\centering
\includegraphics[height=4.0cm,width=4.0cm,angle=0,scale=2.2]{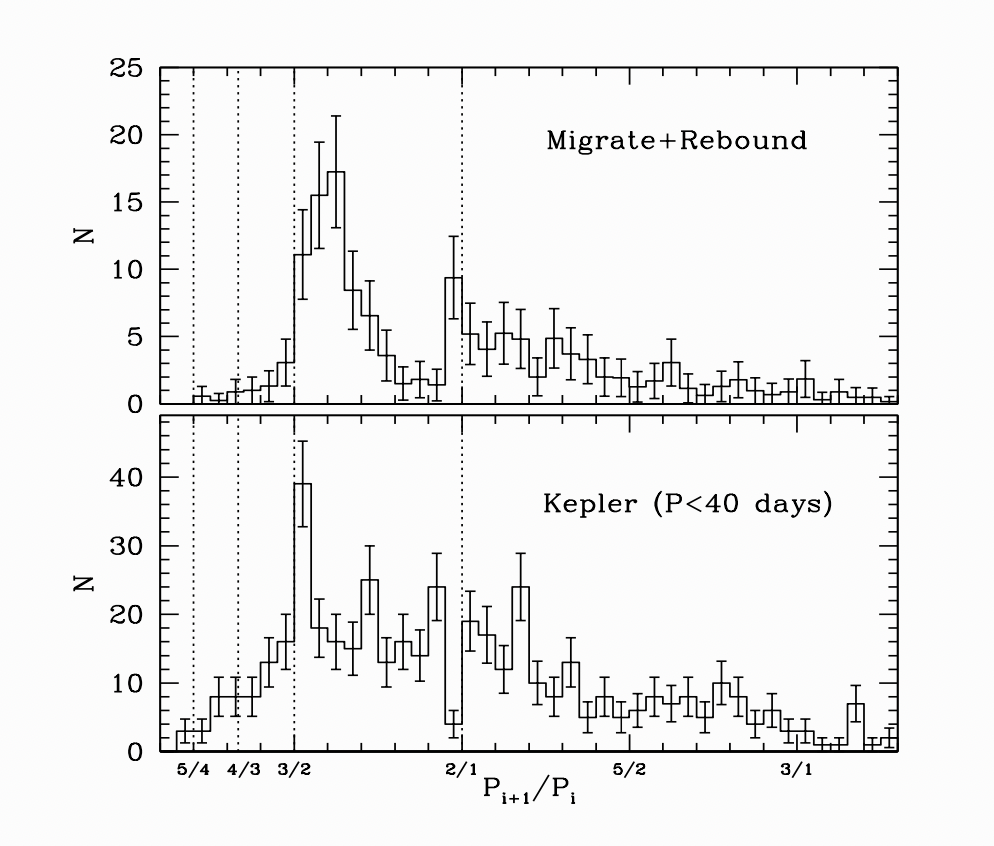}
\caption{The panel shows the distribution of nearest neighbor period ratios from our simulations, when observed by randomly oriented
observers. The lower panel shows the final observed Kepler satellite distribution, when restricted to orbital periods less than 40 days.
Vertical dotted lines indicate the locations of the lowest four first order mean motion resonances.  \label{Pdcomp}}
\end{figure}

As demonstrated in \cite{HYH24}, the final period ratio distribution from this scenario is much more consistent with observations, without invoking
any longer-term dynamical instabilities. 
Figure~\ref{Pdcomp} shows the distribution of nearest neighbour period ratios derived from the full set of dynamical simulations of planetary
systems of multiplicity three or higher.
The ensemble of final configurations
from our full set of simulations (all Kepler analogue mass systems for multiplicity three and above), is 
observed by randomly oriented observers, assuming that all planets that transit are detected (since our masses are derived directly from
the Kepler sample, they are representative of the observed sample). Each system is weighted by the geometric probability of observing the innermost
planet and then additional planets are weighted by the probability of observation relative to the innermost. 
The resulting sample is shown in the upper panel, and includes pairs where one or more intermediate planets are missed because
they do not transit. The lower panel shows the
observed sample of planets, in systems of three or more, from the final Kepler catalog \citep{KeplerFin}, selecting only those planets with orbital periods $<$ 40~days. This is because
our model only covers planets on these scales, and does not contain more distant planets.

The comparison of the upper and lower panels of Figure~\ref{Pdcomp} indicate that the model broadly reproduces the observations,
with a substantial population of pairs between the 3:2 and 2:1 resonances, and a tail extending to larger values (many of these represent
pairs that actually have an nontransitting planet in between them). It is important to note that this emerges directly from the simulations,
and does not require selective weighting of different subsamples of results, as is required to make some other simulations match the data.

\begin{figure}
\centering
\includegraphics[height=4.0cm,width=4.0cm,angle=0,scale=2.2]{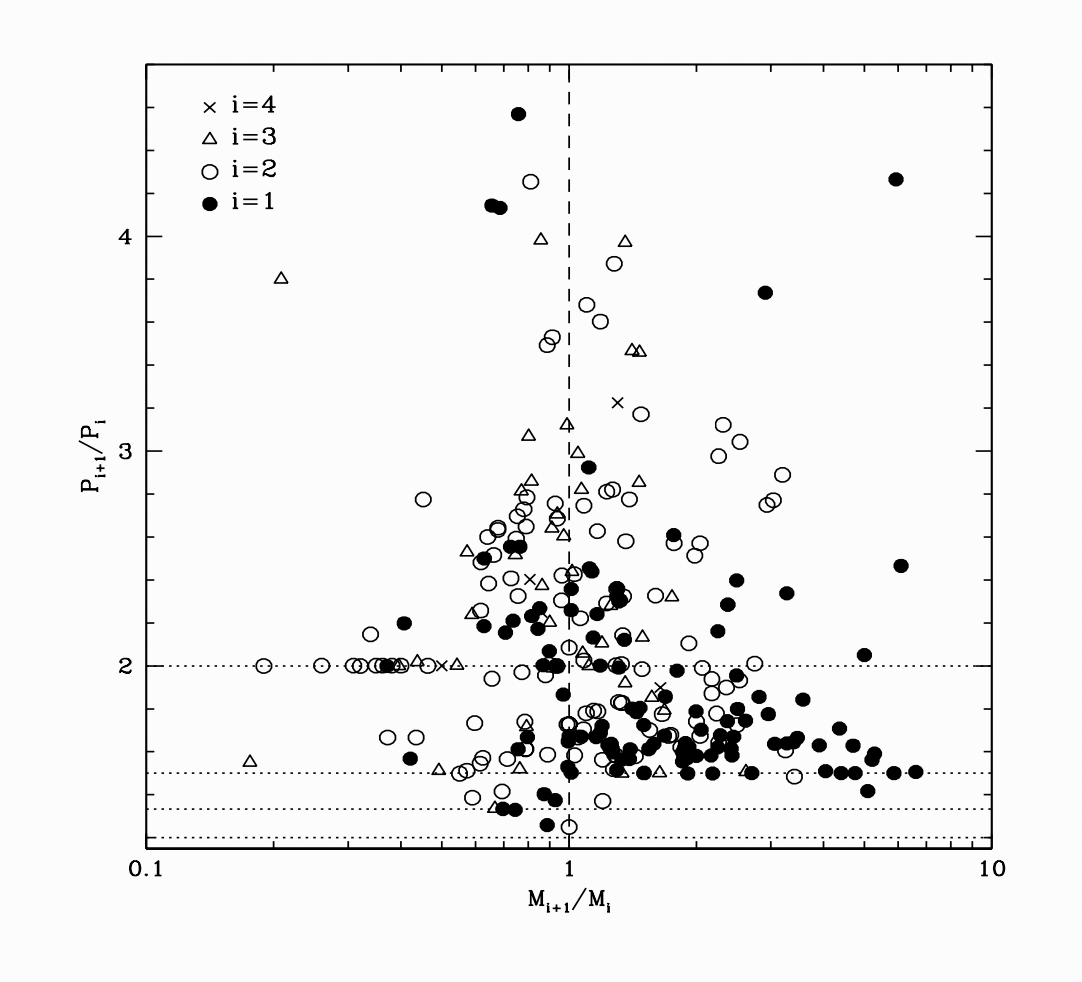}
\caption{The filled circles show the period ratios versus mass ratios for the innermost pairs of multiplanet systems. The open circles
show the ratios for the second and third innermost planets. The open triangles show the ratios for pairs containing the third and
fourth innermost planets. These are fewer because they don't occur in three planet systems. Finally, the crosses indicate the pairs
comprising the fourth and fifth planets in the systems. The horizontal dotted lines indicate the four lowest first order mean motion
resonances and the vertical dashed line indicates mass equailty. \label{MratP}}
\end{figure}

Closer examination of the structure around the 2:1 and 3:2 commensurabilities
indicates that the agreement is not perfect, however. In particular, the model shows a symmetric structure around the 2:1 resonance, whereas
the observations show a clear asymmetry. Similarly, the model peak around the 3:2 resonance is broader than that seen in the observed sample.
 Clearly, while the model does an excellent job of breaking most of the resonant chains, there are still some elements missing.
We will discuss possible causes of this below.

\begin{figure}
\centering
\includegraphics[height=4.0cm,width=4.0cm,angle=0,scale=2.2]{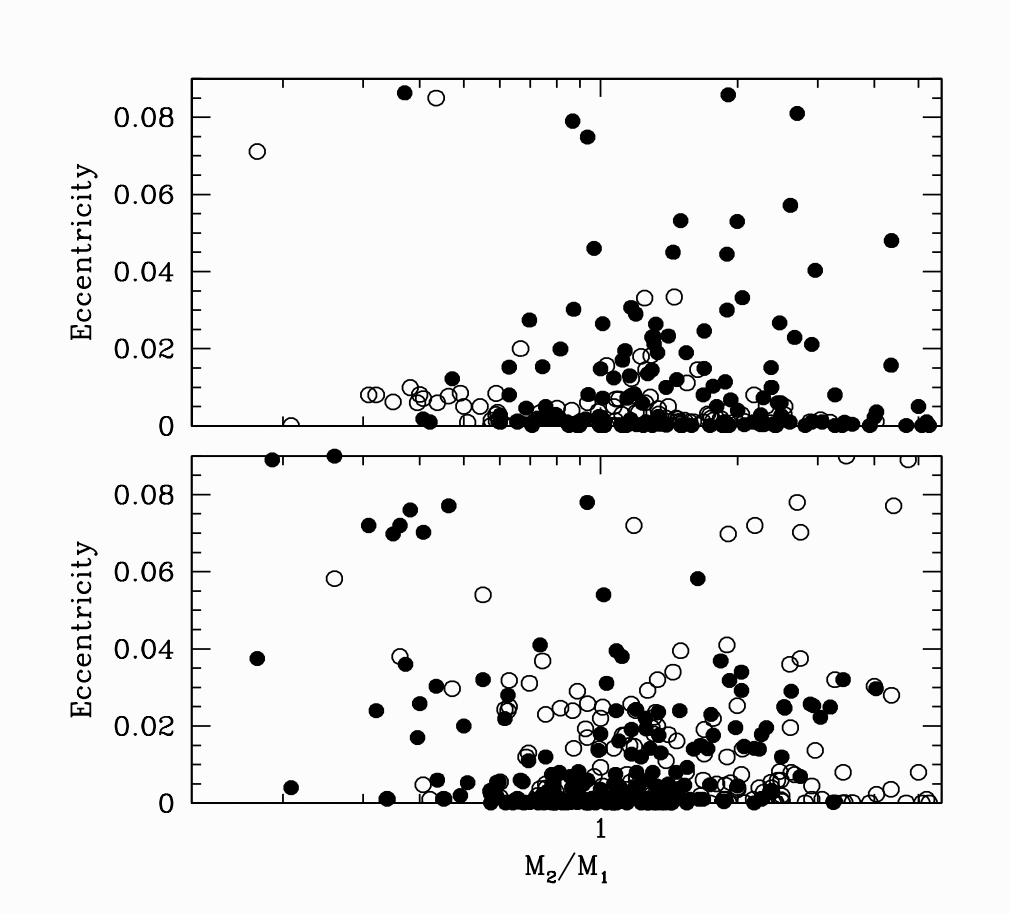}
\caption{The solid points in the upper panel represent the eccentricity of the innermost planet in the higher multiplicity planetary systems (three or more), as a function
of the inner pair mass ratio. The open points in the upper panel show the eccentricity of the outermost planet in each system, along with the mass ratio of the outermost pair. The lower panel shows the eccentricities of all other planets in these high multiplicity systems, against the mass ratio of both pairs to which they belong. Each eccentricity value appears
twice in this plot -- as a solid point when it is the inner member of a pair, and as an open point when it is the outermost member of a pair.  \label{MeMore}}
\end{figure}

As discussed in \cite{HYH24} and \S~\ref{PairSection}, the mass ratio between neighbouring pairs is an important driver of when migration of pairs is divergent or convergent. Figure~\ref{MratP}
shows the period ratios of neighboring pairs versus the mass ratio between the corresponding pairs. In this case, unlike Figure~\ref{Pdcomp}, only true neighbors are
plotted (we do not account for transit probabilities here). Furthermore, the symbols are chosen to represent the position of each pair in the chain, because the presence of planets interior or exterior to a
pair can affect the rate of convergence or divergence (gravitational interactions with these planets can impart an effective inertia).

The distribution of points in Figure~\ref{MratP} does explain  the  peak at the 2:1 resonance.  This arises from pairs in which the inner planet
is significantly more massive than the outer member. During the inward migration phase, the inner planets migrate faster and the pairs
diverge beyond the 2:1 resonance. Once the inner planet reaches the torque reversal, the pair converges again and the pair captures
into the 2:1 resonance. This can be seen in the upper right panel of Figure~\ref{Kep80}. During the rebound phase, the inner planet remains coupled to the disk for longer, thereby maintaining the convergent
migration and resonant lock through to the final decoupling of the planets from the disk.
Our initial mass distribution
was based on the observed Kepler masses, but the constraints were not very strict \citep{CK17}, so that one could reduce the size of the 2:1 peak in Figure~\ref{Pdcomp} 
by restricting the mass ratios from being too small, or by enforcing initial conditions that favor capturing into closer resonances than 2:1. 
The peak at
 the 3:2 resonance, as well as the population just wide of it, is composed largely of inner pairs in which the outer member is more massive.
 In this instance, the outer planet remains coupled to the disk for
longer, leading to divergence from resonance, which explains the large population from 1.5--1.8. Those that remain trapped in the peak are
often slowed down by the presence of additional planets exterior to them. Indeed, the solid points in Figure~\ref{MratP} are clearly biased to
smaller period ratios than the open points, indicating that their divergence is partially impeded by the gravitational interaction with the
planets that lie further out in the planetary system.

As described in \S~\ref{PairSection}, the mass ratio between neighbouring pairs of planets also has an effect on the residual eccentricities. Figure~\ref{MeMore}
shows the equivalent of Figure~\ref{PairMe}, but for pairs that are part of higher multiplicity systems. The upper panel shows the inner planet (solid points) and
outer planet (open points) of each system, as they are only members of a single pair, and so this plot is most analagous to Figure~\ref{PairMe}. The lower panel shows the eccentricities of all the other planets in these systems. The eccentricities of these planets appear once as a solid point,
and once as an open point, as each of these planets is a member of two pairs. 
 We see that
the division into three classes of eccentricity excitation as a function of mass ratio is much less clear here, a consequence of the greater variety of dynamical
evolution that becomes possible with multiple planets interacting with each other and the disk torques.

\subsection{Dynamical Excitation}
\label{Exciting}

\begin{figure}
\centering
\includegraphics[height=4.0cm,width=4.0cm,angle=0,scale=2.2]{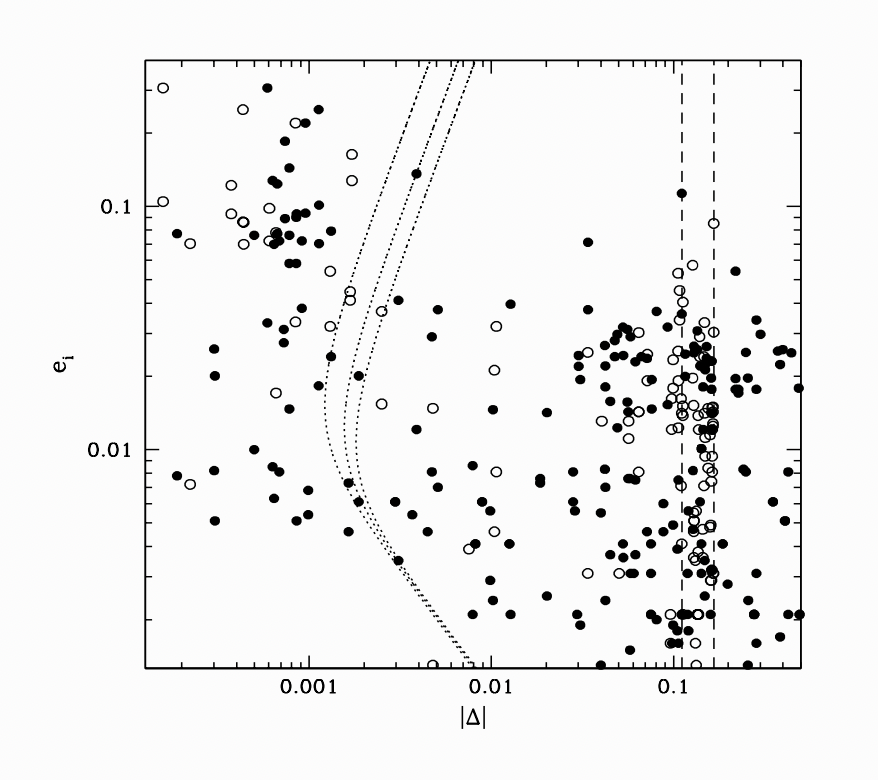}
\caption{The filled circles are for $\Delta >0$ and the open circles are for pairs with $\Delta<0$. The dashed curves indicate the libration amplitudes
for the first order resonances with $q$=1, 2 and 3. The vertical dashed lines indicate the transition between proximity to the q=2 resonance and 
the q=1 resonance. 
 \label{Dele}}
\end{figure}

Our goal is to examine the residual dynamical excitation in our simulations. This is shown in 
  Figure~\ref{Dele}, which 
shows the final eccentricities of each planet in a neighbouring pair in the models presented in Figure~\ref{MratP}. This is 
plotted as a function of $\left| \Delta \right|$ -- the distance to the nearest first order mean motion resonance, given by the commensurability $(q+1)/q$.
This is expressed as 
\begin{equation}
\Delta = \frac{P_{i+1}/P_i}{(1+q)/q} - 1 \label{Delta2}
\end{equation}
 Those for which $\Delta>0$ are shown as a
filled circle and those with $\Delta <0$ as an open circle. The dotted curves indicate the libration amplitudes of the $q=1, 2$ and 3  resonances for a $10 M_{\oplus}$
planet, as given by equation~(8.76) of \cite{MD2000}.  Thus, points to the left of these curves are most likely still trapped in a resonant state, whereas those
to the right are no longer resonant.

The pileup between the two vertical dashed lines is a consequence of the definition of $\Delta$. Members of a pair that diverges from
the 3:2 (q=2) resonance, for example, will move to the right in this diagram as a filled circle, until it reaches the first dashed line. Beyond this point (a period ratio of 5:3) it
is closer to the q=1 commensurability than the q=2, and so jumps to the second dashed line. After this, it moves to the left, as an open circle, until it passes through the q=1 commensurability and again begins
to move to the right as a filled circle.

The points to the left of the dotted lines in Figure~\ref{Dele} demonstrate that the resonant lock of some pairs will survive the late divergent migration, and that these represent the systems with the largest
remnant eccentricities ($\sim 0.1$). These values are imposed during the migration -- a consequence of maintaining resonance lock during migration \citep{TP19}. The
values will evolve as the disk torques decrease, but the resonant libration will maintain them at non-zero levels in the face of the damping by
the protoplanetary disk (which eventually goes away). The population to the right of the dotted lines represent pairs that have experienced divergent
migration away from resonance, so that the eccentricities should have been damped by the protoplanetary disk from their original resonant values. As shown above,
eccentricity pumping during late-time divergent crossing of resonances can populate this area to the upper right in Figure~\ref{Dele}.

\section{Will tides remove residual eccentricity?}
\label{Tides}

The eccentricities shown in Figure~\ref{Dele} represent those at the end of the planetary assembly phase, i.e. they
reflect the outcome of the dynamical interactions which produced the observed planetary system. While it would very
informative to probe these distributions directly, we must account for the fact that these planets lie close to their host
stars and thus the distribution of eccentricities may be substantially modified by tides. We focus here on tidal dissipation
in the planet itself, as our interest in small planets suggests that tides raised in the star will not be important.

A calculation of tidal effects from first principles is quite uncertain as there are many potentially contributing physical
effects \citep{OG14}. A common approach to average over the various possibilities is to adopt a parameterised
tidal strength, most commonly in terms of the `Tidal Q', following \cite{GS66}.  We adopt a version of this model
from \cite{JGB08}, in which the eccentricity of an isolated planet damps as
$e(t)=e(0) exp(-t/\tau)$, where
\begin{equation}
\tau = \frac{4}{63} Q' \frac{M_p a_0^{13/2}}{R_p ^5  (G M_*^3)^{1/2}}
\end{equation}
and where $Q'$ is a measure of the dissipation strength, $a_0$ is the semi-major axis of the planet, $R_p$ and $M_p$ are the planetary radius
and mass, and $M_*$ is the host star mass. When we adopt common values (discussed further below) for these numbers
\begin{equation}
\tau_e = 73 \,  {\rm Gyrs} \left(\frac{Q'}{10^3} \frac{M_p}{M_{\oplus}}\right) \left( \frac{R_{\oplus}}{R_p}\right)^{5} \left( \frac{M_{\odot}}{M_*}\right)^{1/2}
\left( \frac{a_0}{0.1 AU} \right)^{13/2}
\end{equation}
There is still some ambiguity in this expression, as we must specify a mass-radius relation for the planet.

We assume two broad classes of planet in this sample. We assume that all planets with mass $<10 M_{\oplus}$ are rocky, with mass
radius-relation given by the Silicate equation of state from \cite{ZJH21}. For planets with mass $>10 M_{\oplus}$, we assume the
mass-radius relationship from \cite{LF14}, assuming a $1\%$ Hydrogen mass fraction and the high irradiation case.
For the rocky planets, 
we  assume $Q'=10^3$. Although the empirical value of $Q'$ estimated for Earth is significantly smaller,
it is believe that this is dominated by 'pelagic' processes in the oceans, which are likely to be absent in such hot planets as discussed here.
We adopt the larger value of Q', more characteristic of dry, rocky planets, as calibrated by estimates for Mars and the elastic component of Earth's response \citep{L16}.
For the more massive planets, with Hydrogen envelopes, we adopt $Q'=10^4$ for
this, as estimated for Neptune \citep{BM92}.

\begin{figure}
\centering
\includegraphics[height=4.0cm,width=4.0cm,angle=0,scale=2.2]{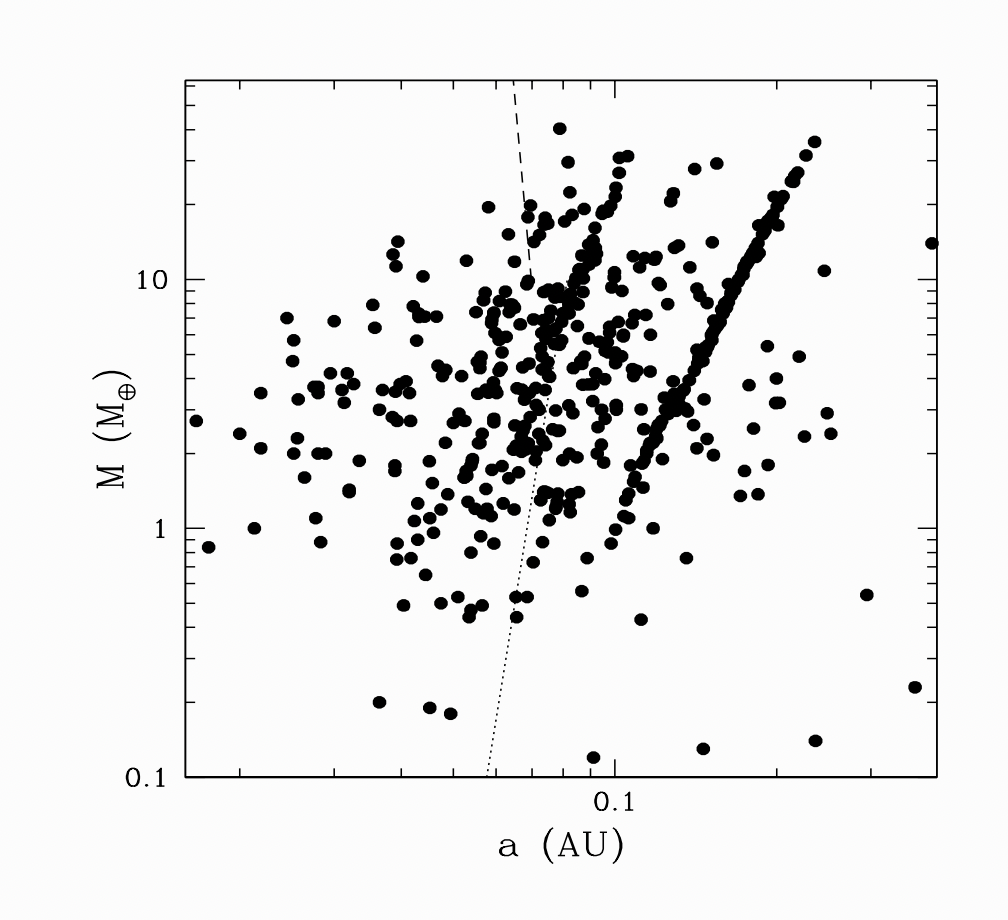}
\caption{The solid points show the locations of the planets at the end of our simulations. The diagonal structure represents the locus of the outermost
planets in the majority of the systems, and represents the mass-dependance of the freeze-out time (and hence the freeze-out location) from the late outward
migration. The dotted curve represents a tidal dissipation time of 1~Gyr, assuming the model in the text with $Q'=10^3$ and a mass radius relation at fixed
(Earth-like) density.  This is assumed to hold for $<10 M_{\oplus}$. The dashed line represents the same criterion, but for Super-Neptune, with a 
radius fixed at $4 R_{\oplus}$ and $Q'=10^4$.  Within this model, the critical period for tidal circularisation is $\sim 8$~days.  \label{TidalReach}}
\end{figure}

With the adoption of this model, we can assess how likely it is that the planets in our sample have had their eccentricities damped by
tides. Figure~\ref{TidalReach} shows the planets from our simulations plotted in semi-major axis and mass, compared to contours for
which $\tau = 5$ Gyr in the two tidal models we consider here. This plot suggests that many of the planets with orbital periods $< 7$~days
may expect to be circularised, while those at longer periods should retain their primordial eccentricities. 

We incorporate the above tidal model into our final eccentricity estimates for the simulated systems.
To zeroth order, each planet will experience an exponential damping of the eccentricity on the timescale $\tau_e$. As demonstrated by the distribution
in Figure~\ref{TidalReach}, the innermost planets in many of these systems will experience circularisation, although the planets exterior to this region
should retain their eccentricities. The matter is further complicated by the fact that many of these planets are 
in close, compact, systems  and will also experience mutual gravitational interactions, which can transfer angular momentum between the planets,
meaning that the eccentricities will evolve for 
 the system
as a whole, either in
the case of secular interactions \citep{AL06,GVL11,HM15,PL19} or in the case of mean motion resonant interactions \citep{LW12,BM13}.

The latter is particularly important as the best information of eccentricities comes from transit time variations, which are most sensitive close to resonance.
We incorporate our tidal evolution model above into the expression from \cite{LW12} for the evolution of the separation $\Delta$ between two neighboring
planets, assuming the tides are dominated by the inner planet.
\begin{equation}
\Delta_{mig}=0.0024 \frac{\gamma} 
{ Q'^{1/3}} \left( \frac{10 M_{\oplus}}{M_1}\right)^{1/3}  \left( \frac{R_1}{R_{\oplus}}\right)^{5/3}
\left( \frac{a_1}{0.1 AU}\right)^{-13/6}  \label{Dmig}
\end{equation} 
and  $\gamma = \left( 2 f_1 \beta (1+ \beta)\right)^{1/3}$,  $\beta = m_2/m_1 \sqrt{a_2/a_1}$, $f_1$ is a constant associated with the particular resonance (given in \cite{LW12}), and $M_1$,$R_1$ are the mass and
radius of the inner planet (with $M_2$ the mass of the outer planet). Proximity to resonance implies a forced eccentricity that persists even after the free
eccentricity as damped, which is then included.

\begin{figure}
\centering
\includegraphics[height=4.0cm,width=4.0cm,angle=0,scale=2.2]{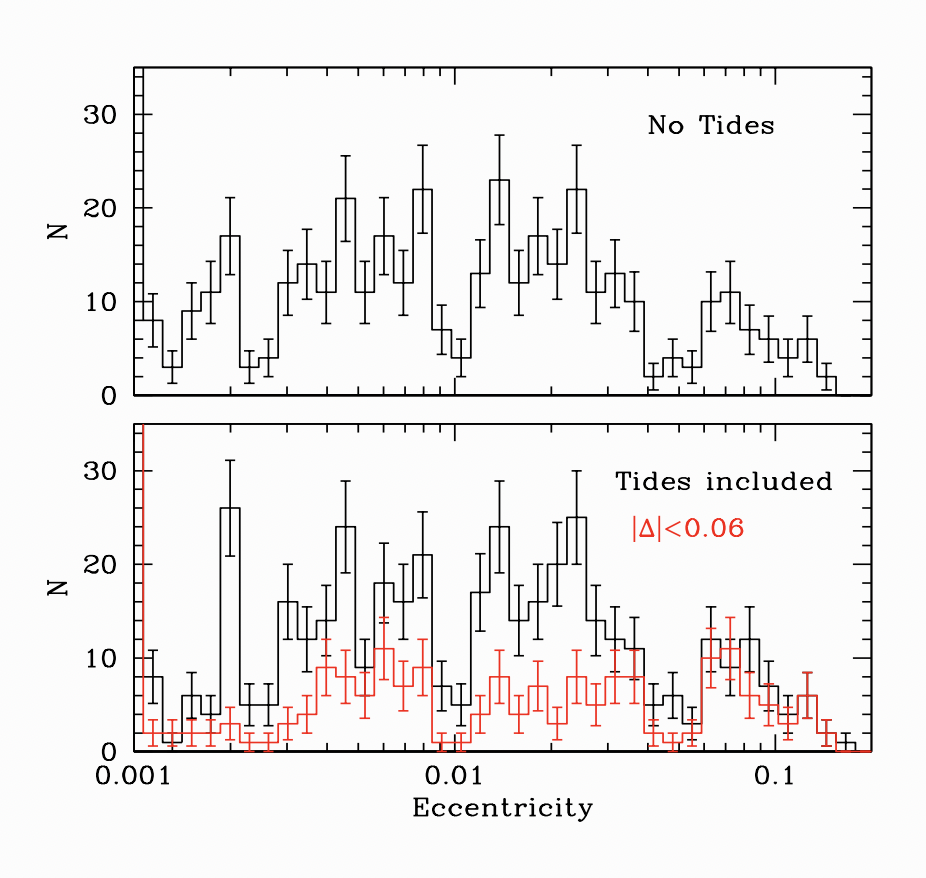}
\caption{The upper panel shows the distribution of eccentricities that emerge from our final population of simulated planetary systems, without any 
correction for tidal effects. The  black histogram in the lower panel shows the same distribution after it has been damped by tides using the model described in the text.
The red histogram shows the same but only for planets that lie in a pair with $\left| \Delta \right|<0.06$.  \label{etides}}
\end{figure}

Figure~\ref{etides} shows the distribution of final planetary eccentricities in our model, both before (upper panel) and after (lower panel) the
incorporation of tidal evolution. While there is a significant population of planets with very low eccentricities, there are also planets with
$e>0.1$. These are mostly systems that remain in resonance even after the dissipation of the gaseous disk. There is also a substantial
population of planets with eccentricities in the range 0.01--0.1, which are well wide of resonance. Overall, $\sim 30\%$ of the simulated
planets have $e>0.01$ (even after correcting for tides).

The red histogram in the lower panel shows eccentricities only for those planets in pairs with $\left| \Delta \right|<0.06$. Much of the information about
observed eccentricities is gleaned from TTV, which favour small $\left| \Delta \right|$. The fraction of the red population that lie $e>0.01$ is 46\%, suggesting that 
measurements from TTV may be biased towards larger eccentricities than the overall population.

\subsection{Inclinations}

Deviations from coplanarity (i.e. inclinations relative to the Laplace plane of the system) can also be a measure of dynamical excitation.
Our simulations were designed to measure evolution in a protoplanetary disk and so initial planetary inclinations were chosen with
a dispersion of only $0.1^{\circ}$. Any significant variation from this would be a sign of dynamical heating.

\begin{figure}
\centering
\includegraphics[height=4.0cm,width=4.0cm,angle=0,scale=2.2]{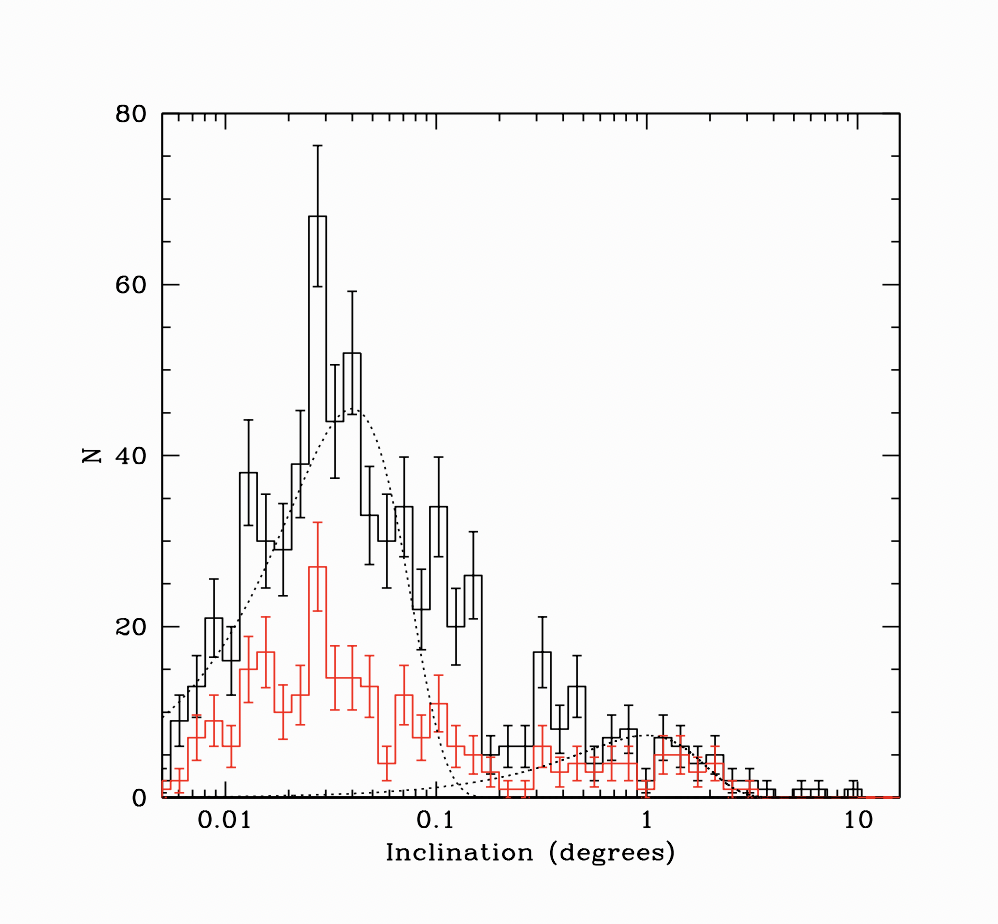}
\caption{The black histogram shows the distribution of final inclinations for our simulated planets. The red histogram
shows the subset which lie in pairs with $\left| \Delta \right| <0.06$.  The two dotted curves indicate Rayleigh distributions
with dispersions of 0.04$^{\circ}$ and 1$^{\circ}$ respectively. \label{ibin2}}
\end{figure}

Figure~\ref{ibin2} shows the distribution of orbital inclinations relative to the plane of the original protoplanetary disk, at the
end of our simulations. The bulk of the distribution remains at inclinations $<0.1^{\circ}$. However, there is a tail that
extends up to $10^{\circ}$. The overall distributions can be roughly fit with two different Rayleigh distributions -- a low
inclination one with dispersion $\sim 0.04^{\circ}$ and a second with a dispersion $\sim 1^{\circ}$, which is similar to
the value inferred from fitting to Kepler multi-planet systems \citep{FM12,XDZ16,VAHM19,MHP19}. The higher inclination group corresponds
to $\sim 16\%$ of the total population and also tend to be found within $\left| \Delta \right|<0.06$,
as can be seen from the comparison of the red and black histograms in Figure~\ref{ibin2}. This comparison  is motivated by suggestions \citep{HFR19} that the observed excess of planets near mean motion resonance
may be a consequence of a reduced inclination dispersion for planets in such configurations. Our results do not
support this hypothesis -- indeed, they go in the other direction. A larger fraction of the low inclination systems are
removed with this $\Delta$ cut.

This higher inclination population are  found in systems in which a pair received sufficient eccentricity excitation
to undergo orbit crossing. Such pairs rapidly become dynamically unstable, resulting
in planetary scattering and eventually collisions.  

\begin{figure}
\centering
\includegraphics[height=4.0cm,width=4.0cm,angle=0,scale=2.2]{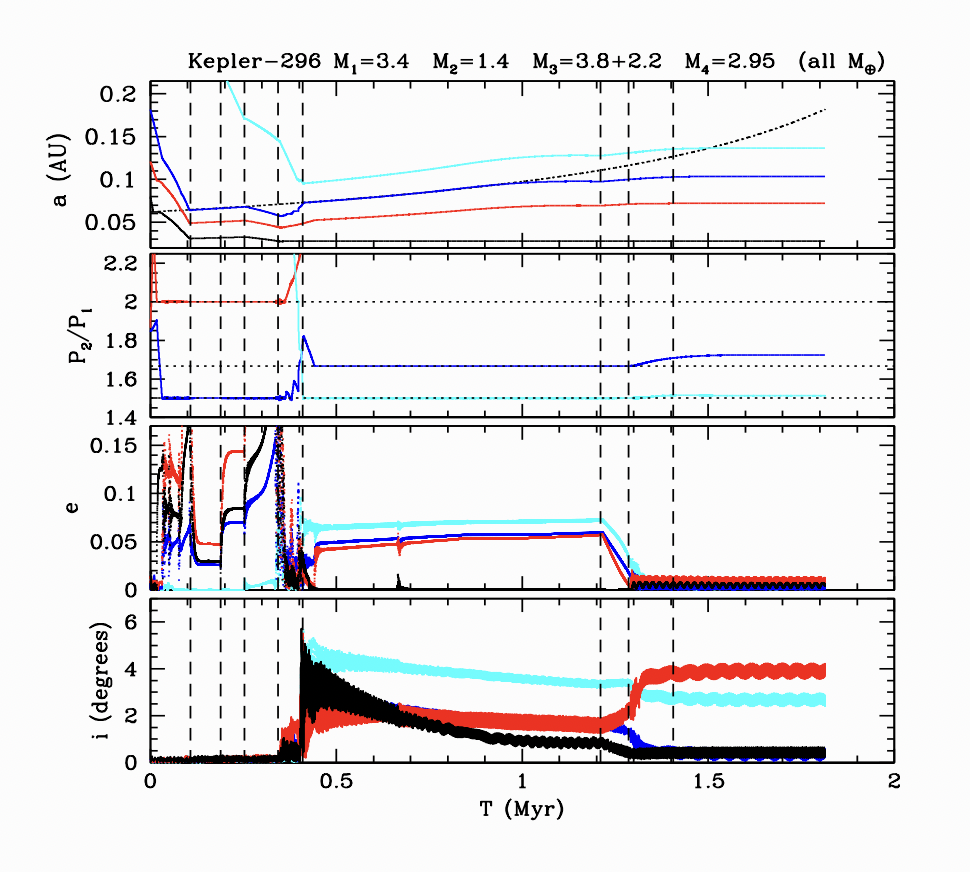}
\caption{The upper panel shows the semi-major axis evolution of the four surviving planets. The
second panel shows the evolution of the period ratios. The third panel shows the evolution of the
eccentricities and the bottom panel shows the evolution of the planetary inclinations. The final
third planet experiences a collision at 0.3 Myr, which generates the first increase in the inclinations. \label{Kep296}}
\end{figure}

An example of such behaviour is shown in Figure~\ref{Kep296}. This shows an analogue of the five planet
system Kepler-296. The system migrates inward as expected, but the third and fourth planets eventually
collide (at 0.3 Myr) which generates an initial spike in the inclination of about 1$^{\circ}$. This is amplified
immediately afterwards, as the system establishes a new resonant chain.
 The 
end result of this shake-out is that the second and third planets move from a 3:2 resonance to a 5:3 second-order
resonance, along with an exterior 3:2 resonance between the third and fourth planets (post collision -- the new third planet is the
sum of the original third and fourth planets).  The fact that the system locks into a second order resonance means that the
inclinations also adjust, as part of the resonance locking.
There is also some late-time adjustment in the inclinations
associated with the breaking of the chains as well.

\section{Comparing to the observations}
\label{Excite}

We have quantified the level of residual dynamical excitation to be expected from the simulated systems in our
simulated planetary systems. In this section, we compare these results to what is known about the observed
compact planetary systems that motivated this model.

\subsection{Free Eccentricities}

The statistical distribution of eccentricities and inclinations of planetary orbits can be measured via the
observation of transit durations, when allied with accurate characterisation of the host star dimensions. 
The application of this technique to the Kepler systems with multiple transitting planets \citep{XDZ16,VAHM19,MHP19,SB23,GPE25}
suggests low eccentricities, with varying analyses placing the mean eccentricity in the range $\sim 0.01$--0.05.

\begin{figure}
\centering
\includegraphics[height=4.0cm,width=4.0cm,angle=0,scale=2.2]{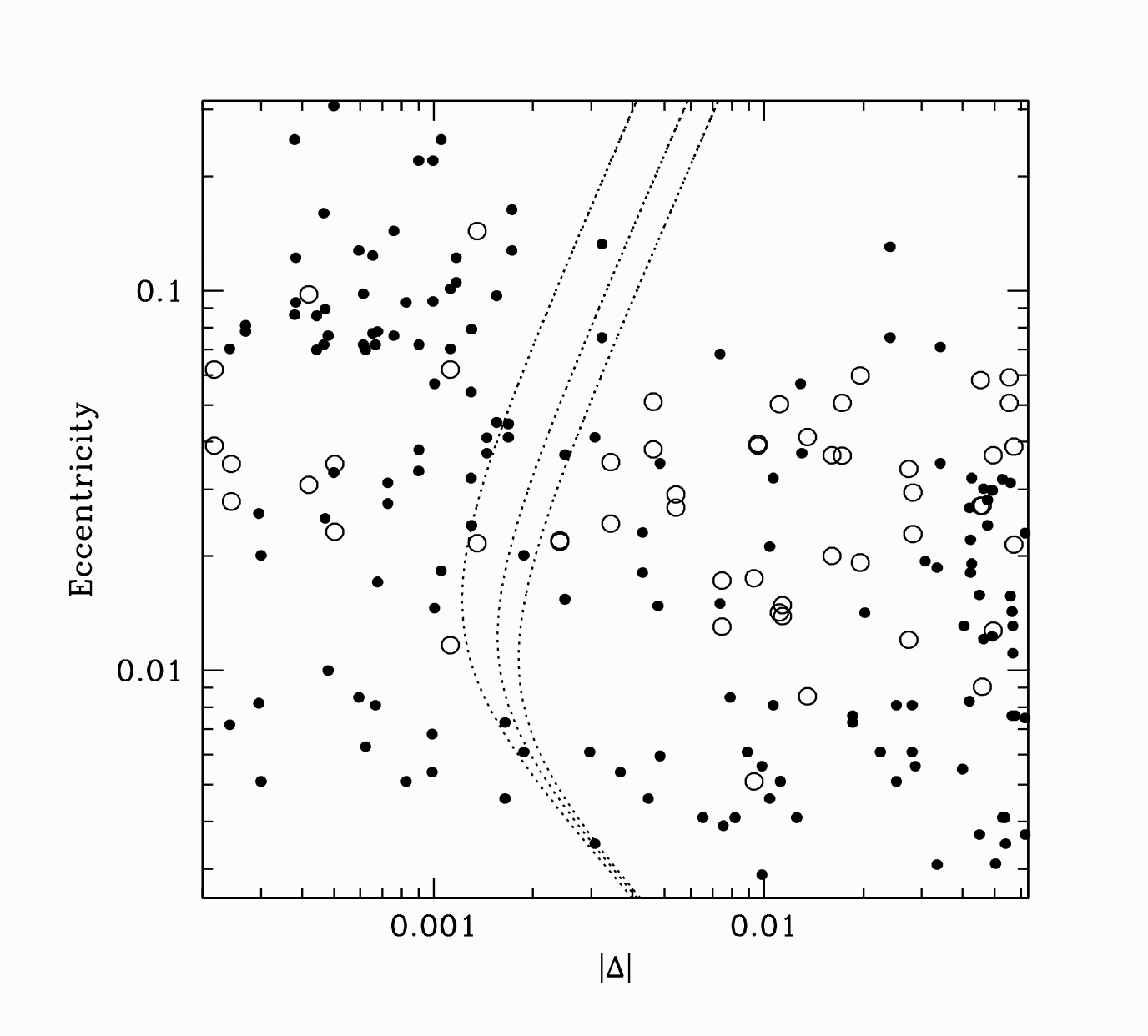}
\caption{The filled circles represent the eccentricities of our simulated sample, after correction for tidal evolution. The large open circles
are the estimated eccentricites for each planet in Table~17 of \cite{JWF21}, which are systems whose posterior measured distributions
are inconsistent with circular. The dotted lines are the same as in Figure~\ref{Dele}, roughly delineating the boundary between
resonant (to the left) and non-resonant (to the right).  \label{JH21}}
\end{figure}

The best constraints on remnant eccentricities come from the observations of
transit timing variations \citep{ASSC05,HM05,FFS12,LXW12,NKB12,MNH13,SFAF13,X14,HL17,JWF21}, wherein the presence
of nearby companions induce a precession of each planetary orbit.
This, in turn, leads to variations in the time of transit, which can be measured.
 This method is most sensitive to pairs that are
close to mean motion resonances, and allows for 
 the detection of small, but non-zero, remnant free eccentricities in a significant sample of transiting systems \citep{HL14,HL17,JWF21}. 
 This provides a direct measure of the level of remnant dynamical excitation, which can be used to test our models.
 Measurement of eccentricity is somewhat degenerate with planetary mass, however, and so we restrict our observational sample
 to the systems in Table~17 of \cite{JWF21}, whose posterior distributions of measured eccentricities are inconsistent with zero.
 Figure~\ref{JH21} shows the central measured values of these systems, compared to the eccentricities shown in Figure~\ref{Dele},
 once again measured as a function of $\left| \Delta \right|$ -- the proximity to the nearest first order commensurability.

The observed points are plotted without error bars because the goal of this comparison is not to compare individual systems, but
to indicate that the remnant eccentricities in our model are of the same order as estimated for the observed samples, and also
match the distribution in the $\Delta$--eccentricity plane. The observational sample also contains many systems with similar
estimated amplitudes whose posterior are also consistent with zero -- and so are not plotted here --  and so the presence of many simulated systems in the
lower half of the plot are also consistent with observations.

\begin{figure}
\centering
\includegraphics[height=4.0cm,width=4.0cm,angle=0,scale=2.2]{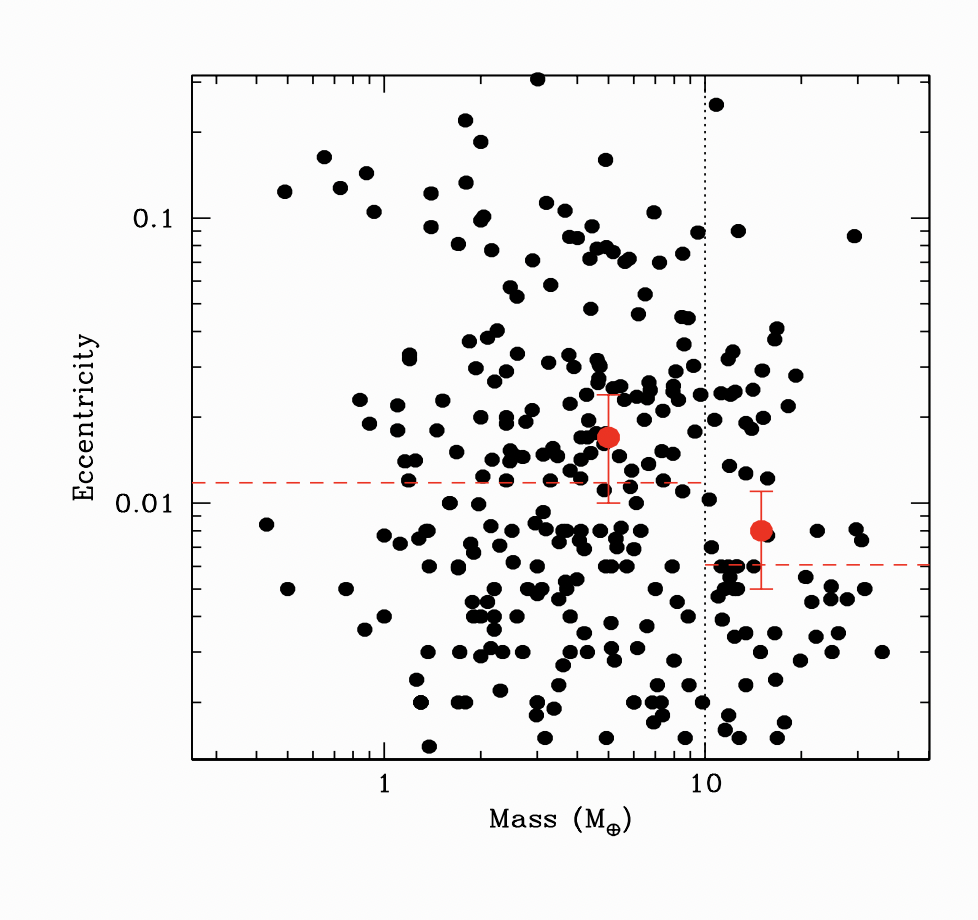}
\caption{The black points show the eccentricity as a function of mass for planets
with finite eccentricity at the end of our simulations. The two red points show the values
inferred by \cite{HL14} for their high mass and low mass populations. We interpret these
as representing populations either above or below the vertical dotted line at 10$ M_{\oplus}$. 
The horizontal red dashed lines represent the median value of the black points
above and below this threshold. \label{Me}}
\end{figure}

Observations have also noted a trend with mass, in the sense that more massive systems have lower 
mean eccentricity \citep{HL14}.
 Figure~\ref{Me} shows the eccentricity as a function of planet mass at the end of our simulations.
\cite{HL14} noted that the eccentricity inferred from TTV was smaller in pairs of planets with mass $> 2.5 R_{\oplus}$
than in planets with radii smaller than this. The precise conversion from radius to mass is uncertain, but we
show (as red points) the approximate eccentricity dispersions inferred by \cite{HL14} as applied to planets
either larger than $10 M_{\oplus}$ or smaller. The results are in agreement with the simulations (it is worth
emphasizing the red points are not fit to the black points, but taken from the observations).

There is also a weak mass trend with proximity to resonance. Those neighbouring pairs with $\left| \Delta \right|<0.01$
have a mean total pair mass of $9.9 \pm 0.7 M_{\oplus}$, and those with $0.01 < \left| \Delta \right|<0.1$ have a
mean total pair mass of $11.4 \pm 0.9 M_{\oplus}$. However, wider pairs, with $\left| \Delta \right|>0.1$, have
a mean total mass of $15.2 \pm 0.8 M_{\oplus}$. This trend is consistent with the consequence that more massive planets
remain coupled to the disk for longer and can therefore diverge further, depending on the mass ratio.
Observations \citep{LDB24} also indicate a trend of increasing mass with increasing $\Delta$.

\subsection{Phase Information}

Finite free eccentricities are also inferred as result of  phase offsets in the mutual timing of the transits. 
For a system where the eccentricity is only due to the resonant forcing, the resonant pair
should exhibit only small amplitude librations about the libration center, and this results in an
expectation of alignment between the transit conjunction and mean transit configuration \citep{LXW12,CC23}.
As the free eccentricity increases, the amplitude of this libration increases and the range of phase
offset grows, before eventually entering circulation.

We have examined the transit timing of pairs in our simulated systems with $0.005 < \Delta < 0.1$, and
calculated the transit phase as described in \cite{CC23}. In each case, we have measured the transit
phase of the pair with and without the presence of other planets in the system, as the simple model used to
calculate the phase \citep{LXW12,CC23} is founded on the two planet model, which can become significantly
more complicated in multiplanet systems where additional sources of precession, and the influence of 
neighbouring resonances, can affect the phase (see \cite{LCC25} for a similar discussion). 
Simulated results can show complex phase evolutions on timescales that are not captured by the limited span
of the observations. As a result,
we classify systems as librating simply if their phases explore a limited range $<90^{\circ}$, and
as circulating if the phases explore a larger range. We have also calculated the phases using only the
properties of the pair in question, and in the case where the pair is embedded within the full planetary
system from which it is drawn.

\begin{figure}
\centering
\includegraphics[height=4.0cm,width=4.0cm,angle=0,scale=2.2]{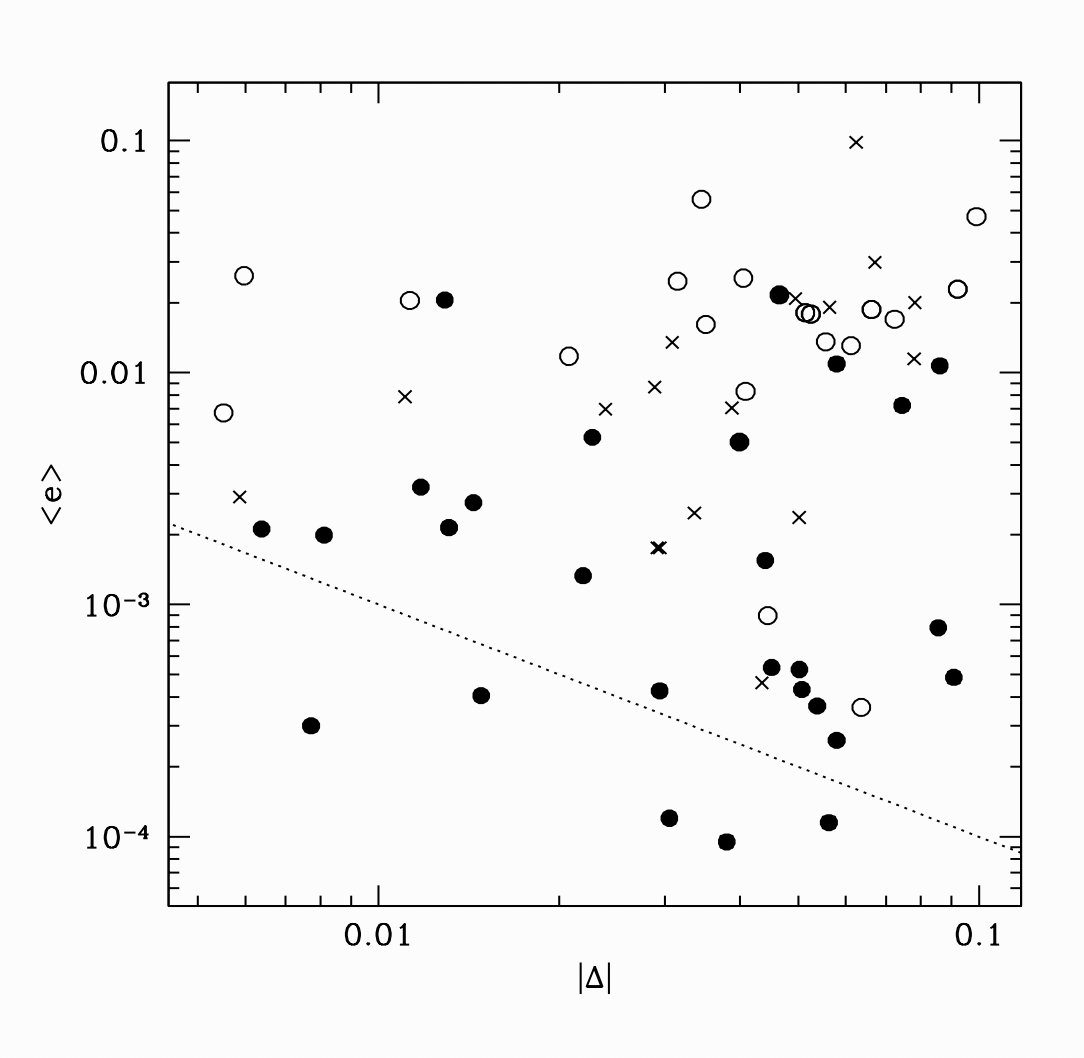}
\caption{We plot each neighbouring pair with a symbol that indicates whether the
TTV phase of the inner member of the pair librates (open symbol), circulates (filled symbol)
or violates the assumptions (crosses)  under which the equations in \cite{CC23} are derived -- which essentially
means that it has not undergone the resonant fixed point bifurcation as discussed in \cite{MD2000}
chapter~8.   \label{Phases}}
\end{figure}

Figure~\ref{Phases} classifies the simulated systems in terms of the behaviour of their transit phases.
We plot $\Delta$
versus mean eccentricity ($<e> = \frac{1}{2} (e_1+e_2)$) for each relevant pair. Systems plotted as solid points exhibit libration
of the transit phase both when we simulate only  the neighbouring pair and also when all the system planets are included.

Open circles show systems that exhibit transit phase circulation, again in both the case of the isolated
pair and with all the planets in the system. The systems plotted as crosses exhibit libration of the 
transit phase for the isolated two planet system, but show phase circulation when the perturbations
from the other planets are included.

Overall, the pattern of points in Figure~\ref{Phases} correspond to expectations.
Systems with $\left| \Delta \right| \sim 0.01$, or larger, still exhibit small amplitude libration if their
eccentricities are small enough. This is the expected behaviour for systems that evolve away
from resonance while eccentricity damping continues to operate. It is also the behaviour expected 
from resonant repulsion as the forced eccentricity
dies away as the pair diverges \citep{LW12,BM13,CC20}. The dotted line in Figure~\ref{Phases} shows the level of eccentricity
consistent with resonance as a function of $\Delta$.

However, if a diverging pair crosses a resonance, it experiences a 
kick to both the eccentricity and the TTV phase. This elevates the system to larger eccentricity
at fixed $\Delta$, moving it towards the upper right in Figure~\ref{Phases}. Indeed, this is where
we find the open points, indicative of circulating $\Phi_{TTV}$.

\subsection{Three Body Resonances}

All of the planetary systems in our simulations form resonant chains, and so they also show libration in
a variety of three-body resonances, as can be seen from the corresponding panels in Figures~\ref{Kep295}--\ref{Kep32}. These panels also show that most of these resonant angles begin circulating as soon as
the two body resonance locks are broken. However, as they diverge, many of the systems remain closer to their three-body commensurabilities than
their two body ones.

In fact, 
\cite{CBG22} find evidence that the Kepler planetary systems show a greater preference for three body commensurabilities than
for two body commensurabilities, which they interpret as potential evidence that three body resonance capture plays an important part
in the migration of compact planetary systems. The weakness of such resonances would therefore require that eccentricity excitation remains
low during the migration process \citep{CMB18,GB21}. Our model provides an alternative route to this excess of proximity to three-body resonance.

\begin{figure}
\centering
\includegraphics[height=4.0cm,width=4.0cm,angle=0,scale=2.2]{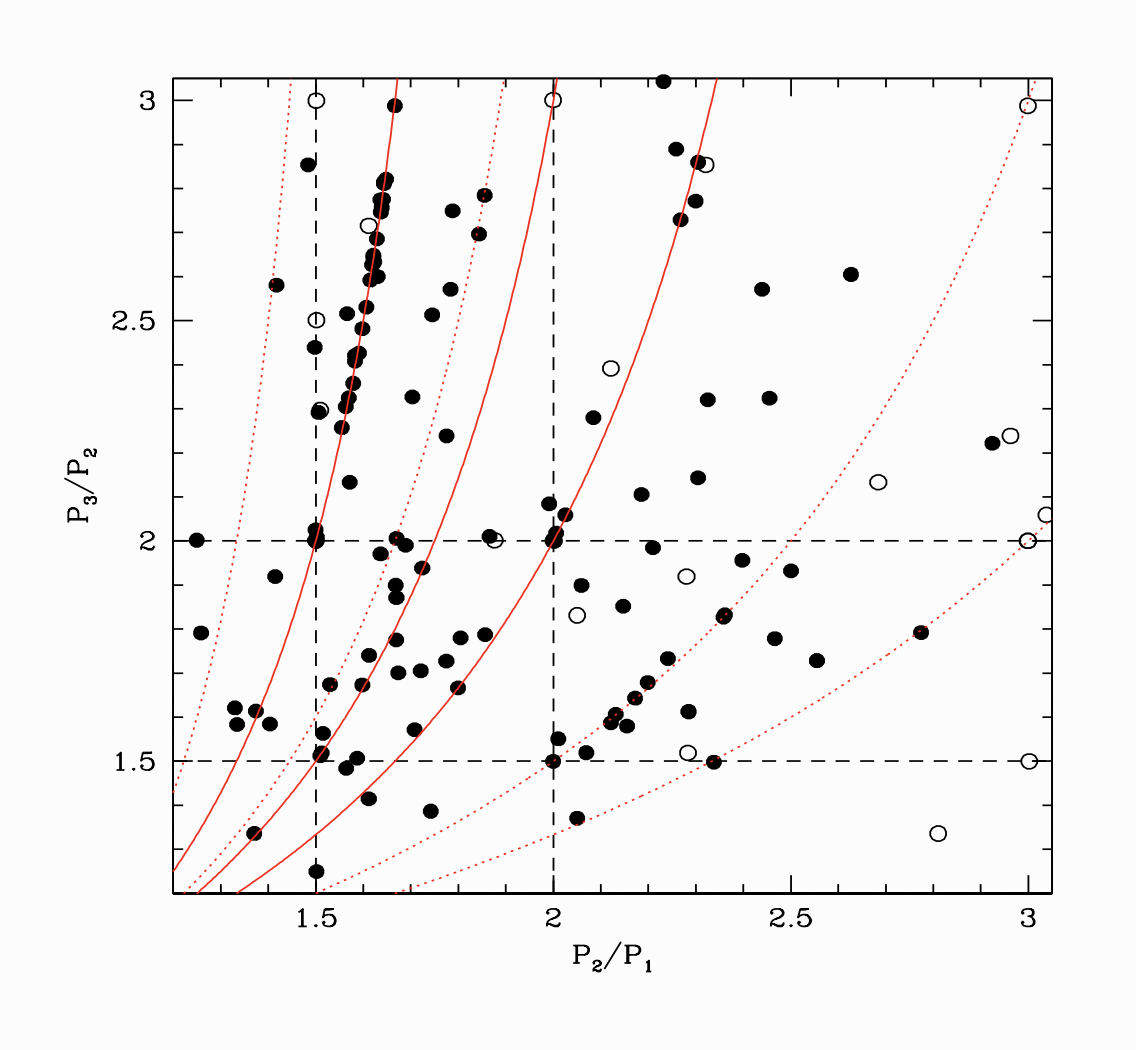}
\caption{The filled circles show the period ratio combinations for the neighbouring triples produced by our simulations. Open
circles show triples produced from a four planet system in which one of the planets is missed (not transitting). There are few
of these shown, because many fall above or to the right of the domain presented here. The red curves represent all of the
three body commensurabilities that can be produced by combining 4:3, 3:2 and 2:1 mean motion resonances. The solid curves
represent the three combinations highlighted by \cite{CBG22}.  The horizontal and vertical dashed lines indicate the common
two body commensurabilities. \label{Web}}
\end{figure}

Figure~\ref{Web}
shows the results from our simulations, composed of all the three-body neighbouring combinations from surviving three, four and 
five body systems.
The red curves represent the possible three body commensurabilities that intersect combinations of the 4:3, 3:2 and 2:1 mean
motion resonances. \cite{CBG22} highlighted three particular combinations, which we show as solid lines. They also restricted their
domain of interest to compact systems, wherein all neighbour period ratios $<1.7$. To compare our results to theirs,  
 we can quantify the
proximity to a three body resonance with
\begin{equation}
\Delta' = k_i \frac{P_2}{P_1} + k_{i+2} \frac{P_2}{P_3} - k_{i+1}.
\end{equation}
Within the domain of interest of \cite{CBG22}, we find that 10\% of our simulations have $\left| \Delta' \right|<0.01$, while the Kepler
sample shows a fraction of 19\% with the same cuts.\footnote{We should note that our observed sample is not exactly the same as
\cite{CBG22}, who drew their sample from an internet archive and thus may exhibit a variety of selection effects.}.
However, if we extend the range of period ratios up to 3, we see that 
 $\sim 40\%$ of the points shown in Figure~\ref{Web} lie within $\left| \Delta' \right|<0.005$ of a three body resonance.
This is artificially high, because it does not yet account for probability of observing a transit. If we model the observability of each system
from random inclinations (and assume it is detected if geometrically possible), then 
 this fraction drops fo 26\% (many of the three-body resonances are wide of the 2:1 MMR, making
them more susceptible to geometric selection effects). However, extending the bounds of the observational sample shows that only
7\% of the observed systems meet this criterion.

Thus, our simulations are consistent with the statistics of \cite{CBG22} for compact systems, but that we see an excess of three-body
commensurabilities in wider spaced systems at the end of our simulations.

\subsection{Dynamical Stabilty}

One important conceptual difference between origin models concerns how early in a planetary system lifetime is the
final dynamical configuration set. The model described here falls into the broad class wherein migration is driven
by interaction with the protoplanetary disk, and the incorporation of a more realistic inner boundary condition results
in a widespread dynamical disruption of resonant chains during the late stages of the protoplanetary disk dispersion.
As a result, the final dynamical configurations  are largely in place at the end of the disk phase (within
a few Myr). Although
further dynamical evolution is not excluded, the comparisons above indicate that the final configurations of the
simulated systems at the end of the disk evolution phase are similar to the observed configurations. 

\begin{figure}
\centering
\includegraphics[height=4.0cm,width=4.0cm,angle=0,scale=2.2]{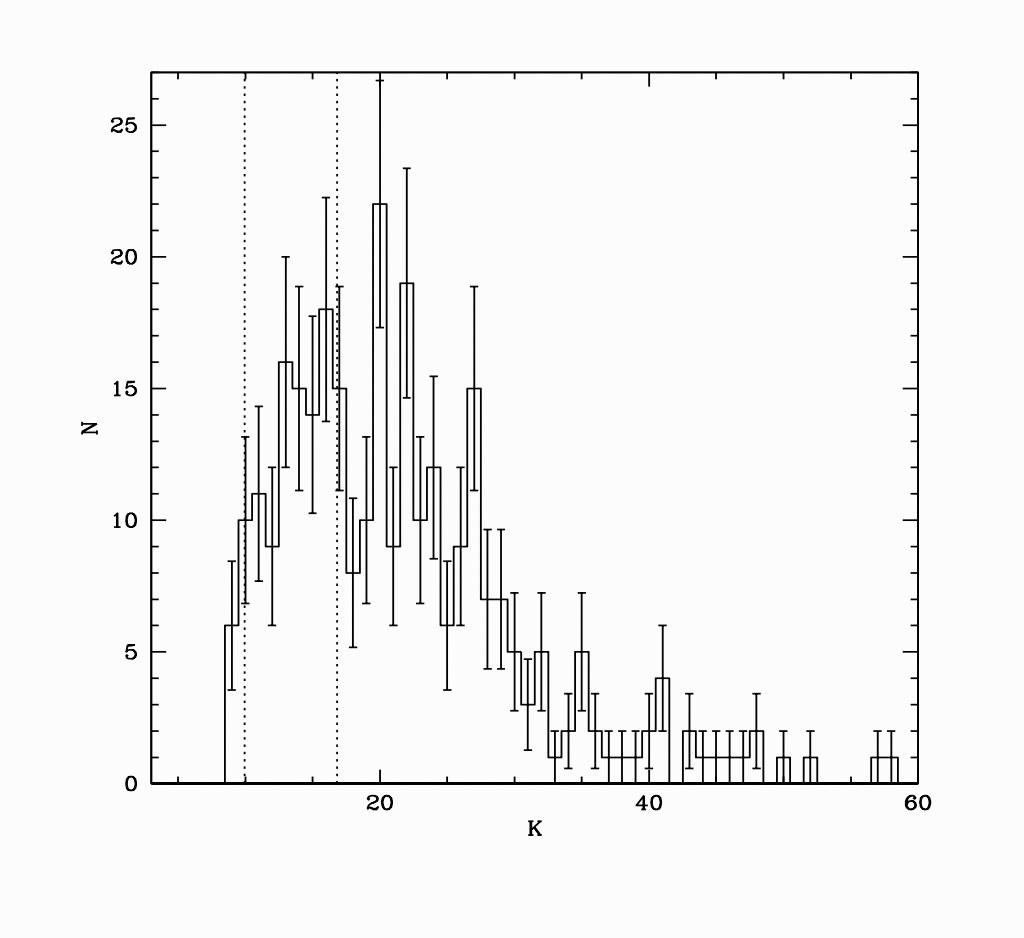}
\caption{The histogram shows the spacing $K$ of neighbouring pairs of planets in the final outputs
of our simulations, expressed in mutual Hill radii. The two vertical dotted lines indicate the value
that would apply for a pair of $10 M_{\oplus}$ planets, in either a 3:2 or a 2:1 mean motion
resonance. \label{Hillbin}}
\end{figure}

The final configurations in our simulations are also broadly consistent with long-term dynamical stability.
An illustration of this is shown in Figure~\ref{Hillbin}, which summarises the final separations of the neighbouring
pairs in our simulations, when measured in terms of mutual Hill radii. We define this separation as
\begin{equation}
K = \frac{(a_{i+1}-a_i)}{(a_{i+1}+a_i)/2} \left( \frac{M_{i+1}+M_{i}}{3}\right)^{-1/3}
\end{equation}
where $a_i$ and $M_i$ are the semi-major axis and mass of planet $i$.  The overall distribution peaks in the
range $15<K<25$, which is well clear of the threshold of short term instability.  The distribution is quite
similar to the empirical determination from \cite{PW15}. The numerical stability simulations of \cite{PW15} also
suggest that the threshold for 50\% stability will lie in the range $K \sim$10--12, for levels of eccentricity
excitation characteristic of our simulations, and for Gyr timescales. This suggests that there is no reason to
expect wholesale dynamical evolution of these systems on longer timescales (absent additional perturbations).

\begin{figure}
\centering
\includegraphics[height=4.0cm,width=4.0cm,angle=0,scale=2.2]{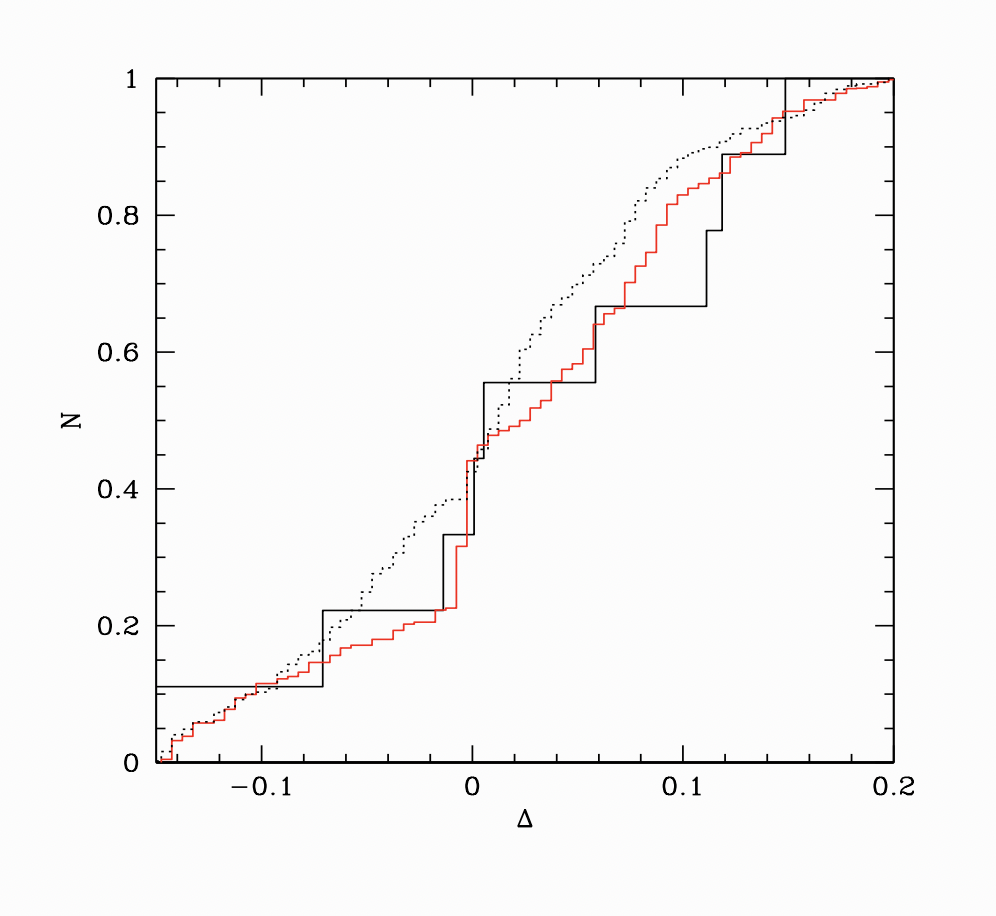}
\caption{The solid histogram shows the distribution of $\Delta$ amongst the  young sample in \cite{DGB24},
edited as discussed in the text.
The red histogram is the corresponding distribution, over the same range in $\Delta$,
from the simulations presented here. The dotted histogram shows the distribution for the
Kepler sample over the same range, subject to the same $P<40$~days cut as in Figure~\ref{Pdcomp}.  \label{Dage}}
\end{figure}

As a result, our models suggest that the dynamical configurations of compact planetary systems are 
largely in place at the end of the protoplanetary disk phase, unlike  other models 
\citep{Izi17,IBR21} that 
 generally   require the systems at the end of
the protoplanetary disk phase  to undergo significant dynamical evolution on longer timescales to disrupt
the resonant chains..

This appears to stand at odds with the recent 
 claim from \cite{DGB24}, namely that the sample of planetary systems orbiting stars with ages $<100$ Myr
has a much higher frequency of near resonant planetary pairs (70$\pm 15$\%) than the overall Kepler sample
($15.6 \pm 1.1$\%). To understand how our simulations compare to this observation, we re-examine 
 the young sample from \cite{DGB24}. Of particular concern is the inclusion of  planets  that
 were initially detected from the Transit Timing Variations they induced in other planets. The amplitudes of TTV are 
 widely known to be stronger for systems close to resonance \citep{ASSC05} and so the inclusion of these
 systems biases the sample in the direction of near-resonant systems. The consequences of this selection effect are
 particular severe for the youngest planet sample in \cite{DGB24}.
 
 Thus, in our comparison, we exclude
 the two young planets (AU Mic~d and TOI-1227c) detected in this way. We also include an additional planet (AU Mic~e)
 that has been detected,  in one of the observed systems, using radial velocities. Finally, \cite{DGB24} cite an unpublished ephemeris for V~1298~Tau,
 so we use that from \cite{FFY22}.
  We also
restrict ourselves to $\Delta$
calculated relative to first order resonances (\cite{DGB24} include second order resonances) to be consistent
with the other comparisons in this paper. With this revised young planet sample, 
 we find 3/9 planetary pairs  satisfy the \cite{DGB24} criterion for proximity to resonance, namely
 $-0.015 < \Delta < 0.03$. This is a fraction of $33 \pm 19$\%, which is less than 1$\sigma$ discrepant from the overall
Kepler rate. If we had included the two planets detected by TTV, this would have risen to 6/11.
The resulting distribution of $\Delta$ for our revision of this young sample is shown as the black solid histogram in Figure~\ref{Dage}, and is compared
with the distribution of $\Delta$ from our simulations (if restricted to the range $-0.15 <\Delta < 0.2$). We see that the shape of the $\Delta$ distribution  around 0 is actually very similar, and closer than the corresponding
distribution from the overall Kepler sample, shown as the dotted histogram. 

Thus, viewed in terms of the distribution around $\Delta \sim 0$, our simulations match those of \cite{DGB24} very well, in accord with new modelling in 
Hu et al. (private communication) which suggests most of this sample is near, but not in, resonance.
There is one important discrepancy though, in the
sense that 
$\sim 44$\% of the simulated neighbouring pairs 
have $\left| \Delta \right| > 0.2$, but there are no such systems in the \cite{DGB24} young stellar sample.
One interpretation of this comparison is that there is some room for later dynamical evolution to broaden the shape of the distribution
around $\Delta \sim 0$ (to move from the red histogram to the dotted histogram in Figure~\ref{Dage})  but that the amount of evolution expected is  a lot less than suggested by \cite{DGB24}.

\section{Conclusions}
\label{Fin}

The residual level of dynamical excitation in compact exoplanet systems is a potentially powerful diagnostic of the 
mechanisms which determine the final state of the planetary system architectures. Here we have examined the level of
excitation to be expected in the scenario outlined in \cite{YHH23} and \cite{HYH24}. The residual eccentricities in
these systems are largely imposed  during the final stage of protoplanetary
disk clearing, as diverging pairs cross mean motion resonances, which results in a transient eccentricity kick to the
system. The fact that most of the divergence occurs at late times also means that the eccentricities are incompletely
damped, leaving a finite value after complete dissipation of the disk.
We have demonstrated that the resulting residual eccentricities are consistent with the level observed in transitting
planet systems, as measured through transit duration distributions and transit timing variations.
We also reproduce the broad trend that more massive
planets are less eccentric. 

We have  shown that the late stage divergence also results in most neighbouring pairs breaking free of resonant
lock, resulting in only a small fraction of planets remaining in resonance. We find that the residual fraction of systems near strict commensurability is similar to that observed, although some details
of the structure near $\Delta \sim 0$ are not reproduced. However, the distribution near $\Delta \sim 0$ is well matched to
that of planetary systems that orbit young stars, although the current sample of the latter population  does appear to lack
the widest systems expected from the simulations.

Our broad conclusion from this comparison is that  a model in which the stellar magnetic field sculpts the inner edge
of the protoplanetary disk can explain the observed dynamical properties of compact extrasolar planet architectures. 
Planetary migration in simple disk models, with a sharp density cutoff at the inner edge, results in long-lived resonant chains, but  the problem is avoided 
if the models include the effect in which the stellar field diffuses
into the disk and modifies the surface density profile. In this case, the magnetospheric rebound at late times leads to widespread divergent
evolution and disruption of the resonant chains. This divergence and disruption also leaves behind a level of dynamical excitation that is
comparable to that observed.

It is natural to compare these results to the currently fashionable paradigm in which the planets emerge from the protoplanetary disk
in resonant chains, which only break up on longer timescales. These models either require an unknown additional source of eccentricity
excitation \citep{LCC25} or capture into very compact configurations of mean motion resonance \citep{IBR21}, which do not correspond to
the observed young planet configurations of \cite{DGB24}. 
In contrast, the magnetospheric rebound model presented here generates the broad features of the dynamical architecture  at the end of protoplanetary disk phase 
without any selective editing to identify specific unstable subsets. As such, it 
does not require widespread dynamical instabilities on longer timescales. This does not preclude some level of additional dynamical evolution, but it
reduces the need considerably, and may lessen the demands on some of the proposed mechanisms. Most importantly, it reproduces the remnant levels
of eccentricity excitation observed through timing transit observations.

{\bf Data availability}: The data underlying this article will be shared on reasonable request to the corresponding author.

The authors thank Nick Choksi \& Fei Dai for comments on a preprint of this work.
Y.H. was supported by the Jet Propulsion Laboratory, California Institute of Technology, under a contract with the National Aeronautics and Space Administration (80NM0018D0004)
This research has made use of the NASA Exoplanet Archive, which is operated by the California Institute of Technology, under contract with the National Aeronautics and Space Administration under the Exoplanet Exploration Program. This research has made use of NASA's Astrophysics Data System Bibliographic Services.

\bibliographystyle{mnras}
\bibliography{refs}

\end{document}